\documentclass[journal]{IEEEtran}

\usepackage{url}
\usepackage{amsmath}
\usepackage{sansmath}
\usepackage{amssymb}
\usepackage{latexsym}
\usepackage{array,arydshln}
\usepackage{multirow}
\usepackage{graphicx}
\usepackage{float}
\usepackage{bm}
\usepackage{tikz}
\usepackage{pgfplots}

\usepackage{ctable}
\usepackage{cite} 

\newtheorem{theorem}{Theorem}

\newtheorem{example}{Example}
\newtheorem{corollary}{Corollary}

\newcommand{\bs}[1]{\ensuremath{\boldsymbol{#1}}}

\newcommand{\CPCCa}{\mathcal{C}_{\mathrm{PCC}}}
 
\newcommand{\CPCCSa}{\mathcal{C}_{\mathrm{SC-PCC}}}
 
\newcommand{\CBCC}{\mathcal{C}_{\mathrm{BCC}}}

\newcommand{\CSCCa}{\mathcal{C}_{\mathrm{SCC}}}
 
\newcommand{\CSCCSa}{\mathcal{C}_{\mathrm{SC-SCC}}}

\newcommand{\CHCCa}{\mathcal{C}_{\mathrm{HCC}}} 
\newcommand{\CHCCSa}{\mathcal{C}_{\mathrm{SC-HCC}}}

\begin{document}
%
\title{Spatially coupled turbo-like codes: \\a new trade-off between waterfall and error floor}



\author{Saeedeh Moloudi, Michael Lentmaier,~\IEEEmembership{Senior Member,~IEEE}, \\ and Alexandre Graell i Amat,~\IEEEmembership{Senior Member,~IEEE}
\thanks{Parts of this paper have been presented at the IEEE International Symposium on Information Theory and Its Applications (ISITA), 2016.}\thanks{This work was supported in part by the Swedish Research Council (VR) under grant \#621-2013-5477.}
\thanks{S. Moloudi and M. Lentmaier are with the Department of Electrical and Information Technology, Lund University, Lund, Sweden (e-mail:  \{saeedeh.moloudi,michael.lentmaier\}@eit.lth.se).}
\thanks{A. Graell i Amat is with the Department of Electrical Engineering, Chalmers University of Technology, SE-41296 Gothenburg, Sweden (e-mail: alexandre.graell@chalmers.se).}
}



\maketitle

\begin{abstract}

Spatially coupled turbo-like codes (SC-TCs) have been shown to have excellent decoding thresholds due to the threshold saturation effect. Furthermore, even for moderate block lengths, simulation results demonstrate very good bit error rate performance (BER) in the waterfall region.
In this paper, we discuss the effect of spatial coupling on the performance of TCs in the finite block-length regime.
We investigate the effect of coupling on the error-floor performance of SC-TCs by establishing conditions under which spatial coupling either preserves or improves the minimum distance of TCs.
This allows us to investigate the error-floor performance of SC-TCs by performing a weight enumerator function (WEF) analysis of the corresponding uncoupled ensembles. 
While uncoupled TC ensembles with close-to-capacity performance exhibit a high error floor, our results show that SC-TCs can simultaneously approach capacity and achieve very low error floor.

\end{abstract}


\IEEEpeerreviewmaketitle

\section{Introduction}

Turbo-like codes (TCs) \cite{BerrouTC} and low-density parity check (LDPC) codes \cite{GallagerLDPCBook} are adopted in many communication standards because they can practically approach the Shannon limit.
Recently, it has been proved that LDPC convolutional codes \cite{JimenezLDPCCC,MichaelCLDPC} ---also known as spatially coupled LDPC (SC-LDPC) codes---exhibit the remarkable threshold saturation phenomenon \cite{Kudekar_ThresholdSaturation,Yedla2012,Yedla2012Vector,Yedla2014}, i.e., for an SC-LDPC ensemble, the belief propagation (BP) decoder can achieve the threshold of the optimal maximum-a-posteriori (MAP) decoder of the underlying uncoupled ensemble.
It then follows that regular SC-LDPC codes achieve capacity as their variable node degrees tend to infinity.
%
%
%
Spatially coupled TCs (SC-TCs) were introduced in \cite{Moloudi_SCTurbo, Moloudi_SCTC_Journal, MoloudiHCC}, and it was proved that threshold saturation also occurs for this class of codes. 
A density evolution analysis shows that, by having stronger component codes, SC-TCs can achieve excellent decoding thresholds with variable nodes of degree one and two only. 

In this paper, motivated by the excellent asymptotic behavior of SC-TCs, we investigate the performance of these codes in the finite block-length regime. 
We consider the same TC ensembles as those in \cite{Moloudi_SCTurbo,Moloudi_SCTC_Journal,MoloudiHCC}, namely parallel concatenated codes (PCCs) \cite{BerrouTC}, serially concatenated codes (SCCs) \cite{Benedetto98Serial,AGiAa}, braided convolutional codes (BCCs) \cite{ZhangBCC,WindowBCC}, and hybrid concatenated codes (HCCs) \cite{3DTurbo, KollerHCC}.
As the first step of our investigation, using the decoding thresholds of the binary erasure channel (BEC) obtained in \cite{Moloudi_SCTC_Journal, MoloudiHCC} and the method described in \cite{LDPCPuncture,UmarISIT2018}, we predict the decoding thresholds over the additive white Gaussian noise (AWGN) channel. Using these thresholds together with the provided simulation results, we discuss the effect of spatial coupling on the performance of TCs in the waterfall region over the AWGN channel.  
Then, we investigate the effect of coupling on the error-floor performance of TCs.
We prove that under certain conditions the minimum distance of a coupled SC-TC ensemble cannot get smaller than that of the corresponding TC ensemble. This means that the error-floor performance of the TCs is not degraded by spatial coupling.
These conditions can be seen as a guideline for unwrapping the TC ensembles.
This connection between the minimum distance of TC and SC-TC ensembles allows us to avoid the complexity of computing the weight enumerator functions (WEFs) of the coupled ensembles. 
Instead, we simply perform a WEF analysis for the uncoupled TC ensemble to investigate and discuss the distance properties of SC-TCs.
Thus, we compute the WEFs of TC ensembles \cite{UnveilingTC,Benedetto98Serial,AbuSurrahGlobeCom07,AbuSurra2011} to obtain bounds on their bit error rate (BER) performance and a bound on the minimum distance. 
Finally, in the last step of our investigation, we use the obtained results to discuss the overall performance of SC-TCs for the finite block-length regime. 

The remainder of the paper is organized as follows.
In Section \ref{Turbo-like Codes}, we briefly describe several TC and SC-TC ensembles by use of the compact graph representation introduced in \cite{Moloudi_SCTC_Journal}.
We discuss the decoding thresholds of these ensembles in Section \ref{Waterfall}.
In the same section, we provide some simulation results to discuss the waterfall region performance of SC-TCs.
In Section \ref{Proof}, we prove that the minimum distance of SC-TC ensembles is either better or equal than that of the corresponding uncoupled ensemble. 
In Section \ref{WE}, we compute the average WEF of TC ensembles to obtain bounds on their BER performance and minimum distance. 
Finally, in Section \ref{con}, we discuss the trade-off between waterfall and error floor performance of SC-TCs, and we conclude the paper in the same section.

\section{Spatially Coupled Turbo-like Codes}\label{Turbo-like Codes}
In this section, we briefly describe four major classes of TCs--- namely, PCCs, SCCs, BCCs, and HCCs--- and their coupled counterparts. 
In particular, we discuss PCCs and SC-PCCs with coupling memory $m=1$,
and refer the interested reader to \cite{Moloudi_SCTC_Journal} for details on the other SC-TC ensembles and higher coupling memories, $m>1$.

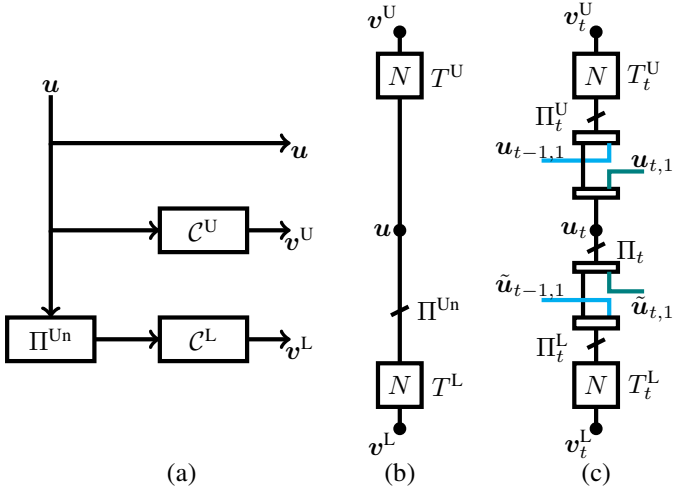
\begin{figure}[t]
	\centering
	
	\begin{tikzpicture} [thick, scale=0.58]
	
	
	\node at (0,3.3) {$\boldsymbol{u}$};
	\draw[->][ultra thick](0,3.1) -- (0,2) -- (5.5,2); 
	\node at (5.7,1.8) {$\boldsymbol{u}$};
	\draw[->][ultra thick](0,2) -- (0,0) -- (2.5,0); 
	\draw[ultra thick] (2.5,-0.5) rectangle (4.5,0.5);
	\node at (3.5,0) {$\mathcal{C}^{\text{U}}$};
	\draw[->][ultra thick](4.5,0) -- (5.5,0); 
	\node at (5.7,-0.2) {$\boldsymbol{v}^{\text{U}}$};
	\draw[->][ultra thick](0,0) -- (0,-2);
	\draw[ultra thick] (-1,-3) rectangle (1,-2);
	\node at (0,-2.5) {$\Pi^{\text{Un}}$};
	\draw[->][ultra thick] (1,-2.5) -- (2.5,-2.5);
	\draw[ultra thick] (2.5,-3) rectangle (4.5,-2);
	\node at (3.5,-2.5) {$\mathcal{C}^{\text{L}}$};
	\draw[->][ultra thick](4.5,-2.5) -- (5.5,-2.5); 
    \node at (5.7,-2.7) {$\boldsymbol{v}^{\text{L}}$};
	\node at (3,-5.6) {(a)};

	\node at (7.6,4.9) {$\boldsymbol{v}^{\text{U}}$};
	\draw [fill=black, ultra thick] (8,4.55) circle [radius=0.1];
	\draw[-][ultra thick](8,4.55) -- (8,4.05); 
	\draw[ultra thick] (7.5,4.05) rectangle (8.5,3.05);
	\node at (8,3.55) {$N$};
	\node at (9.1,3.55) {$T^{\text{U}}$};
	\draw[-][ultra thick](8,3.05) -- (8,0);
	\draw [fill=black, ultra thick] (8,0) circle [radius=0.1];
	\node at (7.6,0) {$\boldsymbol{u}$};
	\draw[-][ultra thick](8,0) -- (8,-3.05); 
	\draw[ultra thick] (7.5,-4.05) rectangle (8.5,-3.05);
	\node at (8,-3.55) {$N$};
	\node at (9.1,-3.55) {$T^{\text{L}}$};
	\draw[-][ultra thick](7.8,-1.95) -- (8.2,-1.75);
	\node at (8.9,-1.8) {$\Pi^{\text{Un}}$};
	\draw[-][ultra thick](8,-4.05) -- (8,-4.55);
	\draw [fill=black, ultra thick] (8,-4.55) circle [radius=0.1];
	\node at (7.6,-4.9) {$\boldsymbol{v}^{\text{L}}$};
	
	\node at (8,-5.6) {(b)};
	
	
	\node at (12.1,4.9) {$\boldsymbol{v}^{\text{U}}_t$};
	\draw [fill=black, ultra thick] (12.5,4.55) circle [radius=0.1];
	\draw[-][ultra thick](12.5,4.55) -- (12.5,4.05); 
	\draw[ultra thick] (12,4.05) rectangle (13,3.05);
	\node at (12.5,3.55) {$N$};
	\node at (13.6,3.55) {$T^{\text{U}}_t$};
	\node at (11.5,2.6) {$\Pi^{\text{U}}_t$};
	\draw[-][ultra thick](12.5,3.05) -- (12.5,2.25);
	\draw[-][ultra thick](12.3,2.5) -- (12.7,2.7);
	\draw[ultra thick] (12,2.25) rectangle (13,2);
	\draw[cyan, ultra thick](11.25,1.6) -- (12.8,1.6)--(12.8,2);
	\node at (11,1.8) {$\boldsymbol{u}_{t-1,1}$};
    \draw[-][ultra thick](12.2,2) -- (12.2,0.95);
	\draw[ultra thick] (12,0.75) rectangle (13,0.95);
	\draw[teal, ultra thick](12.8,0.95) -- (12.8,1.35)--(13.6,1.35);
	\node at (13.8,1.5) {$\boldsymbol{u}_{t,1}$};
	\draw[-][ultra thick](12.5,0.75) -- (12.5,0);
	\draw [fill=black, ultra thick] (12.5,0) circle [radius=0.1];
	\node at (12,0) {$\boldsymbol{u}_t$};
	\node at (13.3,-0.5) {$\Pi_t$};
	\draw[-][ultra thick](12.5,0) -- (12.5,-0.75);
	 \draw[-][ultra thick](12.3,-0.5) -- (12.7,-0.3);
	\draw[ultra thick] (12,-0.75) rectangle (13,-0.95);
	\draw[teal,ultra thick](13.6,-1.4) -- (12.8,-1.4)--(12.8,-0.95);
	\node at (13.8,-1.8) {$\tilde{\boldsymbol{u}}_{t,1}$};
		\draw[-][ultra thick](12.2,-0.95) -- (12.2,-2);
		\draw[cyan, ultra thick](12.8,-2) -- (12.8,-1.6)--(11.25,-1.6);
		\node at (11,-1.3) {$\tilde{\boldsymbol{u}}_{t-1,1}$};
	\draw[ultra thick] (12,-2.25) rectangle (13,-2);
	\draw[-][ultra thick](12.5,-2.25) -- (12.5,-3.05);
		\draw[-][ultra thick](12.3,-2.7) -- (12.7,-2.5);
	\draw[ultra thick] (12,-4.05) rectangle (13,-3.05);
	\node at (12.5,-3.55) {$N$};
	\node at (13.6,-3.55) {$T^{\text{L}}_t$};
	\node at (11.5,-2.7) {$\Pi^{\text{L}}_t$};
	\draw[-][ultra thick](12.5,-4.05) -- (12.5,-4.55);
	\draw [fill=black, ultra thick] (12.5,-4.55) circle [radius=0.1];
	\node at (12.1,-4.9) {$\boldsymbol{v}^{\text{L}}_t$};
	
	\node at (12.5,-5.6) {(c)};
	\end{tikzpicture}
	
	\caption{(a) Encoder block diagram of PCC. Compact graph representation of (b) PCC, (c) SC-PCC.}
	\label{CGPCC}
\end{figure}

Fig.~\ref{CGPCC}(a) shows the block diagram of a rate $R=1/3$ PCC encoder built of two  recursive systematic convolutional encoders, referred to as upper and lower encoder. 
As shown in the figure, the information sequence \bs{u} is encoded by the upper encoder  $\mathcal{C}^{\text{U}}$ to produce the upper parity sequence $\bs{v}^{\text{U}}$.
Likewise, a reordered copy of $\bs{u}$ is encoded by the lower encoder $\mathcal{C}^{\text{L}}$ to produce the lower parity sequence $\bs{v}^{\text{L}}$.
The corresponding permutation is denoted by $\Pi^{\text{Un}}$. 
Finally, the output of the PCC encoder is the sequence $\bs{v}=(\bs{u},\bs{v}^{\text{U}},\bs{v}^{\text{L}})$.

\begin{figure*}[t]
	\centering
	
	\begin{tikzpicture} [thick, scale=0.65]
	

	\node at (0.5,10) {$\boldsymbol{u}$};
	\draw [fill=black, ultra thick] (0.5,9.5) circle [radius=0.1];
	\draw[-][ultra thick](0.5,9.5) -- (0.5,9); 
	\draw[-][ultra thick](0.5,9.5) -- (1.5,9.5); 
	\draw[ultra thick] (0,8) rectangle (1,9);
	\node at (0.5,8.5) {$N$};
	\node at (-0.35,8.5) {$T^{\text{O}}$};
	\draw[-][ultra thick](0.5,8) -- (0.5,7);
	\draw [fill=black, ultra thick] (0.5,7) circle [radius=0.1];
	\node at (0,7) {$\boldsymbol{v}^{\text{O}}$};
	\draw[-][ultra thick](0.5,7) -- (0.5,6.5);
	\draw[-][ultra thick](1.5,9.5) -- (1.5,6.5);
	\draw[ultra thick] (0,6.5) rectangle (2,6.3);
	\draw[-][ultra thick](1,6.3) -- (1,3);
	\draw[-][ultra thick](0.85,4.15) -- (1.15,4.35); 
	\node at (0.4,4.2) {$\Pi^{\text{Un}}$};
	\node at (1.5,3.5) {$\boldsymbol{\tilde{v}}^{\text{O}}$};
	\draw[ultra thick] (0.5,2) rectangle (1.5,3);
	\node at (1,2.5) {$2N$};
	\node at (0.1,2.5) {$T^{\text{I}}$};
	\draw[-][ultra thick](1,2) -- (1,1);
	\draw [fill=black, ultra thick] (1,1) circle [radius=0.1];
	\node at (0.5,1) {$\boldsymbol{v}^{\text{I}}$};
	
	\node at (1,0) {(a)};
	
	
	\node at (4,10) {$\boldsymbol{u}_t$};
	\draw [fill=black, ultra thick] (4,9.5) circle [radius=0.1];
	\draw[-][ultra thick](4,9.5) -- (4,9); 
	\draw[-][ultra thick](4,9.5) -- (5,9.5); 
	\draw[ultra thick] (3.5,8) rectangle (4.5,9);
	\node at (4,8.5) {$N$};
	\node at (3.15,8.5) {$T^{\text{O}}_t$};
	\draw[-][ultra thick](4,8) -- (4,7);
	\draw [fill=black, ultra thick] (4,7) circle [radius=0.1];
	\node at (3.5,7) {$\boldsymbol{v}^{\text{O}}_t$};
	\draw[-][ultra thick](4,7) -- (4,6.5);
	\draw[-][ultra thick](5,9.5) -- (5,6.5);
	\draw[ultra thick] (3.5,6.5) rectangle (5.5,6.3);
	\draw[-][ultra thick](4.35,5.8) -- (4.65,6);
	\node at (5.2,5.9) {$\Pi^{(1)}_t$};
	\draw[-][ultra thick](4.5,6.3) -- (4.5,5.5);
	\node at (6,5.1) {$\tilde{\boldsymbol{v}}_{t,1}$};
	\draw[ultra thick] (3.5,5.5) rectangle (5.5,5.3);
	\draw[teal, ultra thick](5,5.3) -- (5,4.8)--(6,4.8);
	\draw[-][ultra thick](4,4) -- (4,5.3);
	\draw[ultra thick] (3.5,4) rectangle (5.5,3.8);
	\draw[-][ultra thick](4.5,3.8) -- (4.5,3);
	\draw[-][ultra thick](4.35,3.2) -- (4.65,3.5);
	\node at (5.2,3.4) {$\Pi^{(2)}_t$};
	\draw[cyan, ultra thick](5,4) -- (5,4.4)--(3,4.4);
	\node at (2.8,4.25) {$\tilde{\boldsymbol{v}}_{t-1,1}$};  
	\draw[ultra thick] (4,2) rectangle (5,3);
	\node at (4.5,2.5) {$2N$};
	\node at (3.6,2.5) {$T^{\text{I}}_t$};
	\draw[-][ultra thick](4.5,2) -- (4.5,1);
	\draw [fill=black, ultra thick] (4.5,1) circle [radius=0.1];
	\node at (4,1) {$\boldsymbol{v}^{\text{I}}_t$};
	
	\node at (4.5,0) {(b)};
	
	\node at (8.2,10) {$\boldsymbol{v}^{\text{U}}$};
	\draw [fill=black, ultra thick] (8.2,9.5) circle [radius=0.1];
	\draw[-][ultra thick](8.2,9.5) -- (8.2,9); 
	\draw [ultra thick]  (8.4,3) to [out=70,in=0] (8,9.5);
	\draw[-][ultra thick](9.05,8.95) -- (9.35,9.25); 
	\node at (9.4,9.5) {$\Pi^{\text{L}}$};
	\draw[ultra thick] (7.7,8) rectangle (8.7,9);
	\node at (8.2,8.5) {$N$};
	\node at (7.3,8.5) {$T^{\text{U}}$};
	\draw[-][ultra thick](8,8) -- (7.7,5.5);
	\node at (7.2,5.5) {$\boldsymbol{u}$};
	\draw [fill=black, ultra thick] (7.7,5.5) circle [radius=0.1];
	\draw[-][ultra thick](7.7,5.5) -- (8,3); 
	\draw[ultra thick] (7.7,2) rectangle (8.7,3);
	\node at (8.2,2.5) {$N$};
	\node at (7.3,2.5) {$T^{\text{L}}$};
	\draw[-][ultra thick](7.7,4) -- (8.05,4.25);
	\node at (7.2,4) {$\Pi$};
	\draw[-][ultra thick](8.2,1) -- (8.2,2);
	\draw [fill=black, ultra thick] (8.2,1) circle [radius=0.1];
	\draw [ultra thick]  (8.2,1) to [out=0,in=290] (8.5,8) ;
	\draw[-][ultra thick](9.5,2.15) -- (9.8,1.85); 
	\node at (9.6,1.1) {$\Pi^{\text{U}}$};
	\node at (7.7,1) {$\boldsymbol{v}^{\text{L}}$};

	\node at (8.2,0) {(c)};
	
	\node at (12.5,10) {$\boldsymbol{v}^{\text{U}}_t$};
	\draw [fill=black, ultra thick] (12.5,9.5) circle [radius=0.1];
	\draw[-][ultra thick](12.5,9.5) -- (12.5,9); 
	\draw [teal,ultra thick]  (12.5,9.5) to [out=360,in=100] (14,7);
	\draw[-][ultra thick](13.2,8.95) -- (13.55,9.25); 
	\node at (13.6,9.5) {$\Pi^{\text{L}}_t$};
	\draw[ultra thick] (12,8) rectangle (13,9);
	\draw[cyan, ultra thick](12.8,8) -- (13.3,6);
	\node at (13.8,5.9) {$\boldsymbol{v}^{\text{L}}_{t-1}$};
	\node at (12.5,8.5) {$N$};
	\node at (11.5,8.5) {$T^{\text{U}}_t$};
	\draw[-][ultra thick](12.3,8) -- (12,5.5);
	\node at (11.5,5.5) {$\boldsymbol{u}_t$};
	\draw [fill=black, ultra thick] (12,5.5) circle [radius=0.1];
	\draw[-][ultra thick](12,5.5) -- (12.3,3); 
	\draw[cyan, ultra thick](12.8,3) -- (13.3,5);
	\node at (13.8,5.1) {$\boldsymbol{v}^{\text{U}}_{t-1}$};
	\draw[ultra thick] (12,2) rectangle (13,3);
	\node at (12.5,2.5) {$N$};
	\node at (11.5,2.5) {$T^{\text{L}}_t$};
	\draw[-][ultra thick](12,4) -- (12.35,4.25);
	\node at (11.6,4) {$\Pi$};
	\draw[-][ultra thick](12.5,1) -- (12.5,2);
	\draw [fill=black, ultra thick] (12.5,1) circle [radius=0.1];
	\draw [teal,ultra thick]  (12.5,1) to [out=0,in=250] (14,3.5) ;
	\draw[-][ultra thick](13.4,2.15) -- (13.7,1.85); 
	\node at (13.6,1.3) {$\Pi^{\text{U}}$};
	\node at (12,1) {$\boldsymbol{v}^{\text{L}}_t$};
	
	\node at (12.5,0) {(d)};
	\node at (16,10) {$\boldsymbol{u}$};
	\draw [fill=black, ultra thick] (16,9.5) circle [radius=0.1];
	\draw[-][ultra thick](16,9.5) -- (16,9); 
	\draw[-][ultra thick](16,9.5) -- (18,9.5); 
	\draw[-][ultra thick](16.85,9.35) -- (17.15,9.65);
	\node at (17.1,10) {$\Pi^{\text{L}}$};
	\draw[-][ultra thick](18,9.5) -- (18,9);
	\draw[ultra thick] (15.5,8) rectangle (16.5,9);
	\node at (16,8.5) {$N$};
	\node at (15.1,8.5) {$T^{\text{U}}$};
	\draw[ultra thick] (17.5,8) rectangle (18.5,9);
	\node at (18.9,8.5) {$T^{\text{L}}$};
	\node at (18,8.5) {$N$};
	\draw[-][ultra thick](18,8) -- (18,6.5);
	\draw[-][ultra thick](16,8) -- (16,6.5);
	\draw [fill=black, ultra thick] (18,7) circle [radius=0.1];
	\node at (17.5,7) {$\boldsymbol{v}^{\text{L}}$};
	\draw [fill=black, ultra thick] (16,7) circle [radius=0.1];
	\node at (15.5,7) {$\boldsymbol{v}^{\text{U}}$};
	\draw[ultra thick] (15.5,6.5) rectangle (18.5,6.3);
	\draw[-][ultra thick](17,6.3) -- (17,3);
	\draw[-][ultra thick](16.85,4.15) -- (17.15,4.35); 
	\node at (16.4,4.2) {$\Pi^{\text{I}}$};
	\draw[ultra thick] (16.5,2) rectangle (17.5,3);
	\node at (17,2.5) {$2N$};
	\node at (16.1,2.5) {$T^{\text{I}}$};
	\draw[-][ultra thick](17,2) -- (17,1);
	\draw [fill=black, ultra thick] (17,1) circle [radius=0.1];
	\node at (16.5,1) {$\boldsymbol{v}^{\text{I}}$};
	
	\node at (17,0) {(e)};
	
	
	\node at (20.5,10) {$\boldsymbol{u}_t$};
	\draw [fill=black, ultra thick] (20.5,9.5) circle [radius=0.1];
	\draw[-][ultra thick](20.5,9.5) -- (20.5,9); 
	\draw[-][ultra thick](20.5,9.5) -- (22.5,9.5); 
	\draw[-][ultra thick](21.35,9.35) -- (21.65,9.65);
	\node at (21.6,10) {$\Pi^{\text{L}}_t$};
	\draw[ultra thick] (20,8) rectangle (21,9);
	\node at (20.5,8.5) {$N$};
	\node at (19.65,8.4) {$T^{\text{O}}_t$};
	\draw[-][ultra thick](20.5,8) -- (20.5,7);
	\draw [fill=black, ultra thick] (20.5,7) circle [radius=0.1];
	\node at (19.8,7) {$\boldsymbol{v}^{\text{U}}_t$};
	\draw[-][ultra thick](20.5,7) -- (20.5,6.5);
	\draw[-][ultra thick](22.5,9.5) -- (22.5,9);
	\draw[ultra thick] (22,8) rectangle (23,9);
	\node at (22.5,8.5) {$N$};
	\node at (23.5,8.4) {$T^{\text{L}}_t$};
	\draw[-][ultra thick](22.5,8) -- (22.5,6.5);
	\draw [fill=black, ultra thick] (22.5,7) circle [radius=0.1];
	\node at (23.3,7) {$\boldsymbol{v}^{\text{U}}_t$};
	\draw[ultra thick] (20,6.5) rectangle (23,6.3);
	\draw[-][ultra thick](21.35,5.8) -- (21.65,6);
	\node at (22.2,5.9) {$\Pi^{(1)}_t$};
	\draw[-][ultra thick](21.5,6.3) -- (21.5,5.5);
	\draw[ultra thick] (20,5.5) rectangle (23,5.3);
	\draw[teal, ultra thick](22.5,5.3) -- (22.5,4.8)--(23.5,4.8);
	\node at (23.5,5.1) {$\tilde{\boldsymbol{v}}_{t,1}$};
	\draw[-][ultra thick](20.5,4) -- (20.5,5.3);
	\draw[ultra thick] (20,4) rectangle (23,3.8);
	\draw[-][ultra thick](21.5,3.8) -- (21.5,3);
	\draw[-][ultra thick](21.35,3.2) -- (21.65,3.5);
	\node at (22.2,3.4) {$\Pi^{(2)}_t$};
	\node at (21.05,4.95) {$\tilde{\boldsymbol{v}}_{t,0}$}; 
	\draw[cyan, ultra thick](22.5,4) -- (22.5,4.4)--(19.5,4.4);
	\node at (19.5,4.25) {$\tilde{\boldsymbol{v}}_{t-1,1}$}; 
	\draw[ultra thick] (21,2) rectangle (22,3);
	\node at (21.5,2.5) {$2N$};
	\node at (20.6,2.5) {$T^{\text{I}}_t$};
	\draw[-][ultra thick](21.5,2) -- (21.5,1);
	\draw [fill=black, ultra thick] (21.5,1) circle [radius=0.1];
	\node at (21,1) {$\boldsymbol{v}^{\text{I}}_t$};
	
	\node at (21.5,0) {(f)};
	\end{tikzpicture}
	
	\caption{Compact graph representation of (a) SCC (b) SC-SCC, (c) BCC, (d) SC-BCC, (e) HCC (f) SC-HCC.}
	\label{CG}
\end{figure*}
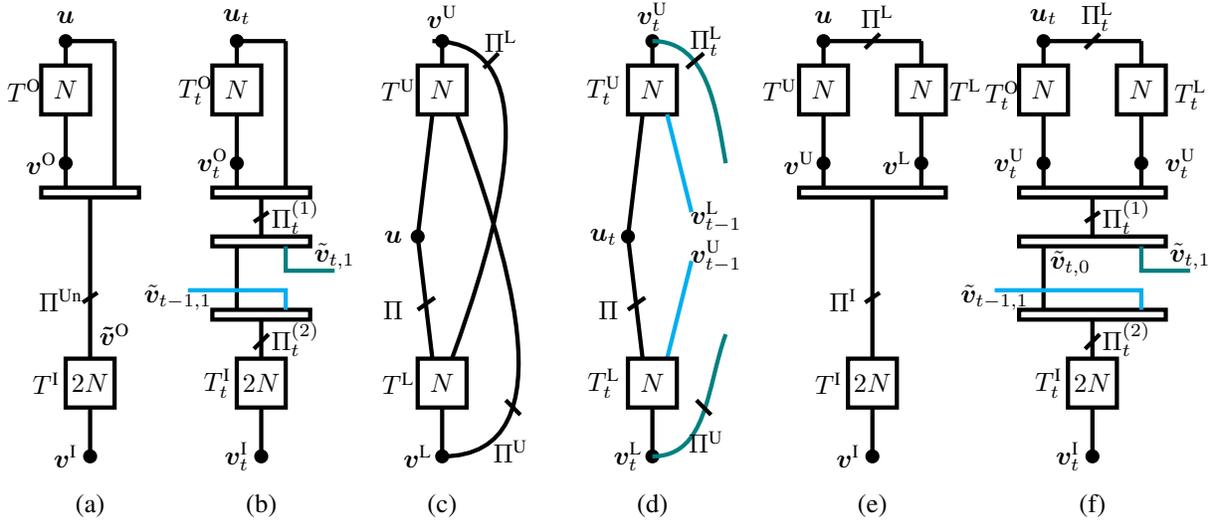


The compact graph representation \cite{Moloudi_SCTC_Journal} of the PCC ensemble is depicted in Fig.~\ref{CGPCC}(b). 
Each of the sequences $\bs{u}$, $\bs{v}^{\text{U}}$, and $\bs{v}^{\text{L}}$ is represented by a black circle, referred to as variable node.
The trellises corresponding to the component encoders are shown by squares, called factor nodes, 
and they are labeled by the length of the trellises. 
The sequences $\bs{u}$ and $\bs{v}^{\text{U}}$ are connected to the upper trellis $\text{T}^{\text{U}}$.
Likewise, a reordered copy of $\bs{u}$ and $\bs{v}^{\text{L}}$ are connected to the lower trellis $\text{T}^{\text{L}}$. 
In order to emphasize that a reordered copy of $\bs{u}$ is connected to $\text{T}^{\text{L}}$, the corresponding permutation is represented by a line that crosses the edge which connects $\bs{u}$ to $\text{T}^{\text{L}}$.

Fig.~\ref{CGPCC}(c) shows the compact graph representation of the spatially coupled PCC (SC-PCC) ensemble with coupling memory $m=1$ at time $t$.
Consider a collection of PCC ensembles at time slots $t=1,\dots,L$, where $L$ is the coupling length.
The SC ensemble can be obtained by dividing the information sequence at time $t$, $\bs{u}_t$, and its reordered copy, $\tilde{\bs{u}}_t$, into two subsequences, denoted by $\bs{u}_{t,j}$ and $\tilde{\bs{u}}_{t,j}$, $j=0,1$, respectively.
Then these subsequences are spread over time $t$ and $t+1$.
The input sequence to the upper encoder at time $t$ is the sequence $(\bs{u}_{t,0},\bs{u}_{t-1,1})$, reordered by permutation $\Pi^{\text{U}}_t$.\footnote{The multiplexer is represented by a rectangular in the compact graph representation.} Likewise, the input sequence to the lower encoder at time $t$ is the sequence $(\tilde{\bs{u}}_{t,0},\tilde{\bs{u}}_{t-1,1})$, reordered by permutation $\Pi^{\text{L}}_t$. 
The information bits at time slots $t\leq 0$ are initialized by zero and the information bits at $t=L$ are chosen in such a way that $\bs{u}_{L,1}=0$ and $\tilde{\bs{u}}_{L,1}=0$ (i.e., we consider the termination of the coupled chain). 

\begin{table*}[t]
	\caption{Predicted AWGN channel thresholds for rate-$1/3$ TCs and SC-TCs.}
	\begin{center}
		\begin{tabular}{lcccccc}
			\hline
			Ensemble&Type&$\text{E}_\text{b}\text{/N}_\text{0}\text{ [dB]}_{\text{BP}}$ & $\text{E}_\text{b}\text{/N}_\text{0}\text{ [dB]}_{\text{MAP}}$  &$\text{E}_\text{b}\text{/N}_\text{0}\text{ [dB]}_{\text{SC}}^1$&$\text{E}_\text{b}\text{/N}_\text{0}\text{ [dB]}_{\text{SC}}^3$ &$\text{E}_\text{b}\text{/N}_\text{0}\text{ [dB]}_{\text{SC}}^5$  \\
			\hline
			$\CPCCa$/$\CPCCSa$&-& -0.1053 & -0.3070& -0.3070 &-0.3070&-0.3070\\[0.5mm]
			$\CSCCa$/$\CSCCSa$&-& 	
			1.4024 &-0.4740 & -0.1196 &-0.4673&-0.4740\\[0.5mm]
			$\CBCC$&Type-I& 1.2139&-0.4723&	
			-0.3992&-0.4573 &-0.4673\\[0.5mm]
			$\CBCC$&Type-II&1.2139 &-0.4723&-0.4690&-0.4723&-0.4723\\[0.5mm]
			$\CHCCa$/$\CHCCSa$&Type-I&	
			3.8846& -0.4941& 1.0366&0.3038&0.0780\\[0.5mm]
			$\CHCCa$/$\CHCCSa$&Type-II& 3.8846&-0.4941&0.2809&-0.4706&-0.4941\\[0.5mm]
			\hline
			
		\end{tabular} 
	\end{center}
	\label{ThresholdsAWGN} 
\end{table*}
%

Fig.~\ref{CG} shows the compact graph representation of the SCC, BCC, and HCC ensembles, and their corresponding spatially coupled ensembles.
In this paper, we restrict ourselves to PCC, SCC and HCC ensembles with identical 4-state component trellises and generator matrix ${\bs{G}}=(1,5/7)$, in octal notation.
For the BCC ensemble, we consider two identical 4-state component trellises with generator matrix
\begin{equation}\label{GMatrix}
\bs{G}(D)= \left( \begin{array}{ccc}1&0&1/7\\0&1&5/7\end{array}\right) \ .
\end{equation}
We also restrict ourselves to systematic TCs and SC-TCs with rate $R=1/3$. 
Therefore, for the SCC and HCC ensembles, we consider full puncturing of the parity sequences of the outer encoders \cite{Moloudi_SCTC_Journal,MoloudiHCC}.  
\section{Spatial Coupling: Waterfall Region Performance                                                                                                                                                                                                                                                                                                                                                                                                                                                                                                                                                                                                                                                                                                                                                                                                                                                                                                                                                                                                                                                    }\label{Waterfall}
\subsection{Asymptotic Performance}
Using the decoding threshold of an ensemble computed for the BEC, it is possible to predict its decoding threshold over the AWGN channel \cite{LDPCPuncture, UmarISIT2018}. 
This allows us to use the decoding thresholds of the TC and SC-TCs from \cite{Moloudi_SCTC_Journal,MoloudiHCC} to predict the corresponding thresholds over the AWGN channel. The results are shown in Table~\ref{ThresholdsAWGN}.
Similar to the BEC, among all the uncoupled TC ensembles, the PCC ensemble has the best BP threshold but the worst MAP threshold.
Conversely, the HCC ensemble has the worst BP threshold but the best MAP threshold, which is very close to the Shannon limit. 
It can also be seen that for all coupled ensembles, threshold saturation occurs.
In general, as the numerical results in Table~\ref{ThresholdsAWGN} suggest, SC-TC ensembles can achieve close-to-capacity BP thresholds.
\subsection{Finite Block-Length Performance}
\begin{figure}[t]
	\centering
	\definecolor{mycolor1}{rgb}{0.00000,0.54000,0.52000}%
	\definecolor{mycolor2}{rgb}{0.63529,0.07843,0.18431}%
	\definecolor{mycolor3}{rgb}{1.00000,0.32000,0.21000}%
	\definecolor{mycolor4}{rgb}{0.00000,0.59000,0.08000}%
	\definecolor{mycolor5}{rgb}{0.00000,0.59000,0.20000}%
	\definecolor{mycolor6}{rgb}{0.60000,0.00000,0.30000}%
	\begin{tikzpicture}
	
	\begin{axis}[%
	width=7cm,
	height=6.1cm,
	at={(0.824in,0.664in)},
	scale only axis,
	xmin=0,
	xmax=4,
	xlabel style={font=\color{white!15!black}},
	xlabel={$\text{E}_\text{b}\text{/N}_\text{0}\;\text{(dB)}$},
	ymode=log,
	ymin=1e-10,
	ymax=1,
	yminorticks=true,
	ylabel style={font=\color{white!15!black}},
	ylabel={BER},
	axis background/.style={fill=white},
	xmajorgrids,
	ymajorgrids,
	yminorgrids,
	legend style={at={(0.51,0.47)}, anchor=south west,font=\footnotesize, legend cell align=left, align=left, draw=white!15!black}
	]
	\addplot [color=mycolor1, dashed, line width=1.5pt, mark=o, mark options={solid, mycolor1}]
	table[row sep=crcr]{%
		0	0.19251982421875\\
		0.25	0.1825353515625\\
		0.5	0.1690681640625\\
		0.75	0.140709765625\\
		1	0.084628515625\\
		1.25	0.02889150390625\\
		1.5	0.0045283203125\\
		1.75	0.0003138671875\\
		2	1.05831473309402e-05\\
		2.25	5.4834799827005e-08\\
	};
	\addlegendentry{SCC, $K=1024$}
	
	\addplot [color=mycolor2, dashed, line width=1.5pt, mark=triangle, mark options={solid, rotate=180, mycolor2}]
	table[row sep=crcr]{%
		0	0.1930427734375\\
		0.25	0.183439086914062\\
		0.5	0.172591967773437\\
		0.75	0.156407446289063\\
		1	0.082587548828125\\
		1.25	0.005832568359375\\
		1.5	7.16798803979498e-05\\
		1.7	1.11575837864919e-09\\
		1.75	0\\
		2	0\\
		2.25	0\\
	};
	\addlegendentry{SCC, $K=4096$}
	
	\addplot [color=mycolor1, line width=1.5pt, mark=o, mark options={solid, mycolor1}]
	table[row sep=crcr]{%
		0	0.090793875933733\\
		0.1	0.0842238670888573\\
		0.2	0.070800686296279\\
		0.3	0.0443002741301971\\
		0.4	0.0171135216932222\\
		0.5	0.00447514303961366\\
		0.6	0.000921980399917401\\
		0.7	8.87203699372661e-05\\
		0.8	8.67343148630292e-06\\
		0.9	4.67760382044464e-07\\
		1	1.79277885886327e-08\\
	};
	\addlegendentry{SC-SCC, $K=1024$}
	
	\addplot [color=mycolor6, line width=1.5pt, mark=triangle, mark options={solid, rotate=180, mycolor6}]
	table[row sep=crcr]{%
		0	0.0701610898820424\\
		0.1	0.017219534475949\\
		0.2	0.00116042537965636\\
		0.3	4.63337261565446e-06\\
		0.4	1.58213250648041e-09\\
	};
	\addlegendentry{SC-SCC, $K=4096$}
	
	\addplot [color=mycolor3, dashed, line width=1.5pt, mark size=4.0pt, mark=asterisk, mark options={solid, mycolor3}]
	table[row sep=crcr]{%
		0	0.05203212890625\\
		0.2	0.02433720703125\\
		0.4	0.00817333984375\\
		0.6	0.00198681640625\\
		0.8	0.00038330078125\\
		1	0.000106796387894998\\
		1.2	4.47352625569078e-05\\
		1.4	2.37740078910576e-05\\
		1.6	1.56578311002218e-05\\
		1.8	1.04147357561604e-05\\
		2	7.11717573897689e-06\\
	};
	\addlegendentry{PCC, $K=1024$}
	
	\addplot [color=mycolor4, dashed, line width=1.5pt, mark size=4.0pt, mark=+, mark options={solid, mycolor4}]
	table[row sep=crcr]{%
		0	0.0440502685546875\\
		0.2	0.008068212890625\\
		0.4	0.0004168701171875\\
		0.8	1.358828163533e-05\\
		1.2	5.32742538102407e-06\\
		1.6	2.68697736671556e-06\\
		1.8	1.92531534414703e-06\\
		2	1.3320173464315e-06\\
	};
	\addlegendentry{PCC, $K=4096$}
	
	\addplot [color=mycolor3, line width=1.5pt, mark size=4.0pt, mark=asterisk, mark options={solid, mycolor3}]
	table[row sep=crcr]{%
		0	0.0263513893235528\\
		0.1	0.0160744685520261\\
		0.2	0.00759753872544846\\
		0.3	0.002830368920624\\
		0.4	0.000939602598852041\\
		0.5	0.000325916746948132\\
		0.6	0.000127166994819744\\
		0.7	5.95716265558421e-05\\
		0.8	3.25112107623318e-05\\
		0.9	2.03514312709966e-05\\
		1	1.36601848259809e-05\\
	};
	\addlegendentry{SC-PCC, $K=1024$}
	
	\addplot [color=mycolor5, line width=1.5pt, mark size=4.0pt, mark=+, mark options={solid, mycolor5}]
	table[row sep=crcr]{%
		0	0.0268895536590004\\
		0.1	0.0145323003980086\\
		0.2	0.00361651258879468\\
		0.3	0.000552839972911081\\
		0.4	0.000117946110718639\\
		0.5	3.74229918134973e-05\\
		0.6	1.62992615630661e-05\\
		0.7	8.36527939383865e-06\\
		0.8	4.92285329399379e-06\\
		0.9	3.31333705357143e-06\\
		1	2.33558347471541e-06\\
	};
	\addlegendentry{SC-PCC, $K=4096$}
	
	\end{axis}
	\end{tikzpicture}%
	\caption{Simulation results for PCC, SC-PCC vs. SCC, SC-SCC, $R=1/3$.}
	\label{SCSCC}
\end{figure}
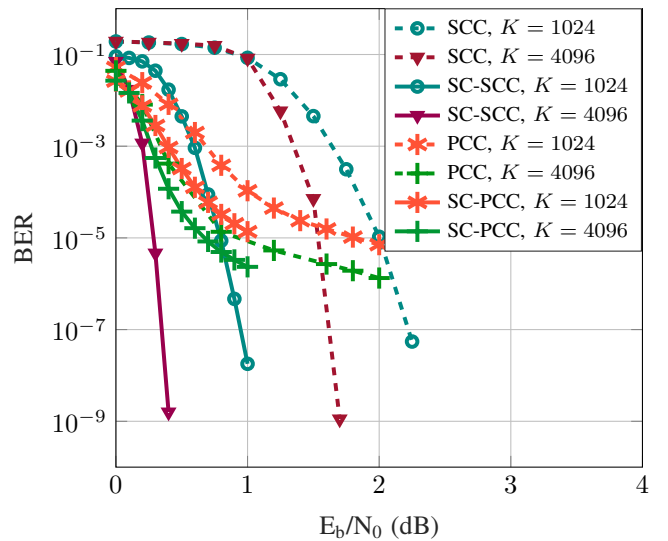

Fig.~\ref{SCSCC} shows BER simulation results for PCCs, SCCs, SC-PCCs, and spatially coupled SCC (SC-SCCs) with $R=1/3$ and input block length $K=1024$ and $K=4096$.
For the coupled ensembles, we consider a coupling length $L=100$ and a sliding window decoder with window size $W=4$ \cite{WindowBCC}. The decoding latency is  $W\cdot K$.
It is well known that the PCC ensemble yields better performance than the SCC ensemble in the waterfall region \cite{Benedetto98Serial}; however, the SCC ensemble has a much lower error floor than the PCC ensemble. 
By applying spatial coupling, the performance of the PCC and SCC ensembles improves significantly for both input block lengths. 
This improvement is more substantial for the SCC ensemble than for the PCC ensemble.
For instance, the performance of the SCC ensemble with $K=1024$ at BER$=10^{-5}$ improves more than $1$ dB with coupling. 
The coupling gains are in agreement with the decoding thresholds in Table~\ref{ThresholdsAWGN}. As it can be seen, the gap between the BP and MAP threshold of the SCC ensemble is larger than that of the PCC ensemble, hence the expected gain from coupling is bigger for the SCC ensemble.

In Fig.~\ref{SCSCC}, the uncoupled ensemble with $K=4096$ and the coupled ensemble with $K=1024$ have equal latency, i.e., both ensembles have a decoding latency of 4096 bits.
For this latency, the SC-SCC ensemble performs better than the SCC ensemble.
However, in the case of PCCs, for a latency of 4096 bits, the uncoupled ensemble performs slightly better than the corresponding coupled ensemble. Interestingly, for equal latency, the SC-SCC ensemble outperforms the SC-PCC ensemble in the waterfall region.
Thus, the SC-SCC ensemble yields better performance in both the waterfall and error floor regions.

In the following section, we investigate the impact of spatial coupling on the error floor performance of TCs.


\section{Spatial Coupling: Error Floor Region Performance}\label{Proof}
Similar to uncoupled TC ensembles, to analyze the performance of SC-TC ensembles in the error floor region, one could derive bounds based on the WEFs of the ensembles.
Unfortunately, deriving the WEF for SC-TCs is cumbersome.
In this section, we establish a connection between the WEF of SC-TC ensembles and that of the corresponding uncoupled ensembles. In particular, we prove that, under certain conditions, spatial coupling does not decrease the minimum distance of TCs.
This allows us to use the WEF analysis of TCs to estimate the error floor performance of SC-TCs.
A similar connection between LDPC and SC-LDPC codes is proved in \cite{TruhachevDistBoundsTBLDPCCCs,MitchellMinDisTrapSet,PseudocodLDPC}.
Here, we restrict ourselves to SC-TCs with coupling memory $m=1$, but the proof can be generalized to higher coupling memories.

\begin{theorem} 
	Consider an uncoupled PCC, $\tilde{\mathcal{C}}$, (see Fig.~\ref{CGPCC}(b)) with permutation $\Pi^{\text{Un}}$
	and parity-check matrices $\bs{H}_{\text{U}}$ and $\bs{H}_{\text{L}}$ corresponding to the upper and lower component encoders.
	It is possible to unwrap the PCC to form an SC-PCC, $\mathcal{C}$ (Fig.~\ref{CGPCC}(c)).
	For the SC-PCC, we assume a length-$L$ coupled chain with termination or tailbiting, and time-invariant permutations. Let us denote the permutations by $\Pi^{\text{U}}_t=\Pi^{\text{U}}$, $\Pi^{\text{L}}_t=\Pi^{\text{L}}$, and $\Pi_t=\Pi$, and assume that they satisfy \[\Pi^{\text{Un}}=(\Pi^{\text{U}})^{-1}\cdot \Pi \cdot \Pi^{\text{L}}.\]
	Then, for any codeword $\bs{v}\in\mathcal{C}$, $\bs{v}=(\bs{v}_1,\dots,\bs{v}_L)$, $\bs{v}_t=(\bs{u}_t,\bs{v}_t^{\text{U}}, \bs{v}_t^{\text{L}})$, with Hamming weight $w_{\text{H}}(\bs{v})$, there exists a codeword $\tilde{\bs{v}}\in\tilde{\mathcal{C}}$ such that 
	\[
	w_{\text{H}}(\tilde{\bs{v}})\leq w_{\text{H}}(\bs{v})\; .
	\]
\end{theorem}

\begin{IEEEproof}
	We prove this theorem for tailbiting of the coupled chain, which contains termination as a special case. The result is thus valid for both cases.  
	Any codeword $\bs{v}\in\mathcal{C}$ satisfies the local constraints for $t=1,\dots,L$. Therefore, at time $t$,
	
	\begin{align}
	\label{PCCProof1}
	&\big((\bs{u}_{t,0},\bs{u}_{t-1,1})\cdot\Pi^{\text{U}}\;\;\; \bs{v}^{\text{U}}_t\big)\cdot \bs{H}_{\text{U}}^{\text{T}}=\bs{0}, \\
	\label{PCCProof2}
	&\big((\bs{u}'_{t,0},\bs{u}'_{t-1,1})\cdot\Pi^{\text{L}}\;\;\; \bs{v}^{\text{L}}_t\big)\cdot \bs{H}_{\text{L}}^{\text{T}}=\bs{0},
	\end{align} 
	where $\bs{u}'_t=\bs{u}_t\cdot \Pi$. The constraints are linear and time-invariant. Thus, for $t={1,\dots,L}$, any superposition of the vectors $\big((\bs{u}_{t-1,1},\bs{u}_{t,2})\cdot\Pi_1\;\;\; \bs{v}^{\text{U}}_t\big)$ and $		\big((\bs{u}'_{t-1,1},\bs{u}'_{t,2})\cdot\Pi_1\;\;\; \bs{v}^{\text{L}}_t\big)$ satisfies \eqref{PCCProof1} and \eqref{PCCProof2}, respectively. In particular, consider
	\begin{align}
	&\sum_{t=1}^{L}\big((\bs{u}_{t,0},\bs{u}_{t-1,1})\cdot\Pi^{\text{U}}\;\;\; \bs{v}^{\text{U}}_t\big)\nonumber \\ &=\big(\sum_{t=1}^{L}(\bs{u}_{t,0},\bs{u}_{t-1,1})\cdot\Pi^{\text{U}}\;\;\; \sum_{t=1}^{L}\bs{v}^{\text{U}}_t\big)\nonumber\\
	&=\big(\sum_{t=1}^{L}\bs{u}_{t}\cdot\Pi^{\text{U}}\;\;\; \sum_{t=1}^{L}\bs{v}^{\text{U}}_t\big)\;,
	\label{PCCProof3}
	\end{align}
	and
	\begin{align}
	&\sum_{t=1}^{L}\big((\bs{u}'_{t,0},\bs{u}'_{t-1,1})\cdot\Pi^{\text{L}}\;\;\; \bs{v}^{\text{L}}_t\big)\nonumber \\ &=\big(\sum_{t=1}^{L}(\bs{u}'_{t,0},\bs{u}'_{t-1,1})\cdot\Pi^{\text{L}}\;\;\; \sum_{t=1}^{L}\bs{v}^{\text{L}}_t\big)\nonumber\\
	&=\big(\sum_{t=1}^{L}\bs{u}'_{t}\cdot\Pi^{\text{L}}\;\;\; \sum_{t=1}^{L}\bs{v}^{\text{L}}_t\big)\nonumber\\
	&=\big(\sum_{t=1}^{L}\bs{u}_t\cdot\Pi\cdot\Pi^{\text{L}}\;\;\; \sum_{t=1}^{L}\bs{v}^{\text{L}}_t\big)\;.
	\label{PCCProof4}
	\end{align}
	
	Let
	
	\[
	\tilde{\bs{u}}=\sum_{t=1}^{L}\bs{u}_t\cdot\Pi^{\text{U}},\;\;\;\tilde{\bs{v}}^{\text{U}}=\sum_{t=1}^{L}\bs{v}_t^{\text{U}},\;\;\;\tilde{\bs{v}}^{\text{L}}=\sum_{t=1}^{L}\bs{v}_t^{\text{L}}\;.
	\]
	Then, the vectors obtained from \eqref{PCCProof3} and \eqref{PCCProof4} can be rewritten as $(\tilde{\bs{u}}\;\;\tilde{\bs{v}}^{\text{U}})$ and $(\tilde{\bs{u}}\cdot\Pi^{\text{Un}}\;\;\tilde{\bs{v}}^{\text{L}})$, respectively.
	
	The vectors from \eqref{PCCProof3} and \eqref{PCCProof4} satisfy \eqref{PCCProof1} and \eqref{PCCProof2}, respectively.
	Thus,
	
	\begin{align}
	&(\tilde{\bs{u}}\;\;\tilde{\bs{v}}^{\text{U}})\cdot \bs{H}_{\text{U}}^{\text{T}}=\bs{0}, \\
	&(\tilde{\bs{u}}\cdot\Pi^{\text{Un}}\;\;\tilde{\bs{v}}^{\text{L}})\cdot \bs{H}_{\text{L}}^{\text{T}}=\bs{0}.
	\end{align} 
	
	Therefore, $\tilde{\bs{v}}=(\tilde{\bs{u}},\tilde{\bs{v}}^{\text{U}},\tilde{\bs{v}}^{\text{L}})$ is a codeword of the uncoupled ensemble.
	
	If all nonzero elements of $\bs{v}_t$, $t=1,\dots,L$, occur at different positions, then  $w_H(\tilde{\bs{v}})=w_H(\bs{v})$. Otherwise, the overlap of the nonzero elements reduces the weight of $\tilde{\bs{v}}$ and $w_{\text{H}}(\tilde{\bs{v}})< w_{\text{H}}(\bs{v})$.
	\end{IEEEproof}

This theorem can be extended to the other TC ensembles.
\begin{theorem}
Consider an uncoupled SCC (BCC/HCC), $\tilde{\mathcal{C}}$, (Fig.~\ref{CG}).
It is possible to unwrap the SCC (BCC/HCC) to form an SC-SCC (BCC/SC-HCC), $\mathcal{C}$ (Fig.~\ref{CG}).
For the coupled code, we assume a length-$L$ coupled chain with termination or tailbiting, and time-invariant permutations which satisfy certain conditions. Then, for any codeword $\bs{v}\in\mathcal{C}$, $\bs{v}=(\bs{v}_1,\dots,\bs{v}_t,\dots,\bs{v}_L)$, $\bs{v}_t=(\bs{u}_t,\bs{v}_t^{\text{U}}, \bs{v}_t^{\text{L}})$, there exists a codeword $\tilde{\bs{v}}\in\tilde{\mathcal{C}}$ such that 
\[
w_{\text{H}}(\tilde{\bs{v}})\leq w_{\text{H}}(\bs{v})\;.
\]	
\end{theorem}
	\begin{IEEEproof}
		See Appendix.
	\end{IEEEproof}
\begin{corollary}
	The minimum distance of an SC-TC ensemble $\mathcal{C}$ is larger than
	or equal to the minimum distance of the underlying uncoupled TC ensemble
	$\tilde{\mathcal{C}}$,
	\[
	d_{\text{min}}(\mathcal{C}) \geq d_{\text{min}}(\tilde{\mathcal{C}}) \; .
	\] \hfill $\square$
\end{corollary}
By the above theorems, we establish conditions on the permutations under which SC-TCs have equal or better minimum distance than their corresponding TCs.
These conditions can be considered as guidelines for selecting proper permutations for SC-TCs.
According to the theorems and the corollary above, the WEF analysis of uncoupled TC ensembles can be used to investigate the error floor and the minimum distance of SC-TC ensembles. 

\section{Weight Enumerator Analysis} \label{WE} 
In this section, we describe how to derive upper bounds on the error rate performance of TC ensembles and bounds on the minimum distance of these ensembles based on their WEFs \cite{UnveilingTC, Benedetto98Serial}. Then, we compare these bounds for different classes of TCs.
For that, we first derive the average input-parity WEF (IP-WEF) of the component encoders.
In particular, we describe the steps for a rate-$2/3$ recursive systematic convolutional encoder. A similar method can be used to derive the IP-WEF of any convolutional encoder with arbitrary rate $R$. 
Then, we use the obtained IP-WEFs to compute the average IP-WEFs of the TC ensembles.

\subsection{Input-Parity Weight Enumerator}\label{WEF}

Let $A(I_1,I_2,P)$ denote the IP-WEF of a rate-$2/3$ recursive systematic convolutional encoder, 
\[
A(I_1,I_2,P)=\sum_{i_1} \sum_{i_2}\sum_{p} A_{i_1,i_2,p} \, I^{i_1} I^{i_2}P^p,
\]
where the coefficient $A_{i_1,i_2,p}$ denotes the number of codewords with weight $i_1$, $i_2$, and $p$
for the first input, the second input, and the parity sequence, respectively.

$A(I_1,I_2,P)$ can be computed as follows.
For a trellis with $s$ states, transitions within a trellis section can be described by an $s\times s$ matrix $\bs{M}$.
The element of $\bs{M}$ in the $r$th row and the $c$th
column, $[\bs{M}]_{r,c}$, corresponds to the trellis branch from the $r$th state to the $c$th state. 
More precisely, $[\bs{M}]_{r,c}$ is a monomial $I_1^{i_1}I_2^{i_2}P^{p}$, where $i_1$, $i_2$, and $p$ are the weights corresponding to the first, second, and third outputs of the transition from the $r$th state to the $c$th state.
For a trellis with $N$ sections, the overall transition matrix is $\bs{M}^N$.
Considering that the trellis is initialized and terminated to the all-zero state, the IP-WEF is given by the element $[\bs{M}^{N}]_{1,1}$.  
\begin{example}
	Assume a terminated, rate-$2/3$ convolutional encoder with three trellis sections and generator matrix in \eqref{GMatrix}.

The transition matrix can be written as
\begin{equation}
\label{eq:AR23}
\boldsymbol{M}(I_1,I_2,P)=\left( \begin{array}{cccc}1&I_2P&I_1I_2&I_1P\\I_1&I_1I_2P&I_2&P\\I_2P&1&I_1P&I_1I_2\\I_1I_2P&I_1&P&I_2 \end{array}\right),\nonumber
\end{equation}
and the IP-WEF becomes
\begin{align}
&A(I_1,I_2,P)= [\bs{M}^{3}]_{1,1}=\nonumber\\
&1+I_2^3P^2+2I_1I_2P+I_1I_2P^3+2I_1I_2^2P+\nonumber\\
&I_1I_2^2P^3+I_1^2I_2+2I_1^2I_2P^2+3I_1^2I_2^2P^2+I_1^3P+I_1^3I_2^3P \;.\nonumber
\end{align}
	\end{example} \hfill $\triangle$

For a rate-$1/2$ convolutional encoder, we can obtain the transition matrix $\bs{M}$ in a similar way.
Then, the IP-WEF of the encoder is given by
$[\bs{M}^{N}]_{1,1}$ and can be written as 
\[
A(I,P)=\sum_{i} \sum_{p}A_{i,p} \,  I^iP^p,
\]
where $A_{i,p}$ is the number of codewords of input weight $i$ and
parity weight $p$.

Consider the PCC ensemble shown in Fig.~\ref{CG}(b).
Let $A^{\text{T}_{\text{U}}}(I,P)$ and $A^{\text{T}_{\text{L}}}(I,P)$ denote the IP-WEFs corresponding to the upper and lower component encoder, respectively.
The overall IP-WEF depends on the IP-WEF of the component encoders and the permutation used.
Averaging over all possible permutations, the coefficients of the average IP-WEF of the PCC ensemble, $\bar{A}_{i,p}^{\text{PCC}}$, can be obtained as \cite{UnveilingTC}
 
\begin{equation}
\bar{A}_{i,p}^{\text{PCC}}=\frac{\sum_{p_1}A^{\text{T}_{\text{U}}}_{i,p_1}\cdot A^{\text{T}_{\text{L}}}_{i,p-p_1}}{\binom{N}{i}}\;.
\end{equation}

For the SCC ensemble shown in Fig.~\ref{CG}(b), we denote the IP-WEFs of the outer and inner encoder by $A^{\text{T}_{\text{O}}}(I,P)$ and $A^{\text{T}_{\text{I}}}(I,P)$, respectively.
Similar to PCCs, the average IP-WEF of the SCC ensemble, $\bar{A}_{i,p}^{\text{SCC}}$, can be computed by averaging over all possible permutations \cite{Benedetto98Serial}. The coefficients $\bar{A}_{i,p}^{\text{SCC}}$ can be written as
\begin{equation}
\label{IPWESCC}
\bar{A}_{i,p}^{\text{SCC}}=\sum_{p_1}\frac{A^{\text{T}_{\text{O}}}_{i,p_1}\cdot A^{\text{T}_{\text{I}}}_{i+p_1,p-p_1}}{\binom{2N}{i+p_1}}\;.
\end{equation}

We denote the IP-WEFs corresponding to the upper and lower component encoders of the BCC ensemble (Fig.~\ref{CG}(c)) by $A^{\text{T}_{\text{U}}}(I,P)$ and $A^{\text{T}_{\text{L}}}(I,P)$, respectively.
The coefficients of the average IP-WEF, $\bar{A}_{i,p}^{\text{BCC}}$, can be computed as

\begin{equation}
\label{IPWEBCC}
\bar{A}_{i,p}^{\text{BCC}}=\sum_{p_1}\frac{A^{\text{T}_{\text{U}}}_{i,p_1,p-p_1}\cdot A^{\text{T}_{\text{L}}}_{i,p-p_1,p_1}}{\binom{N}{i}\binom{N}{p_1}\binom{N}{p-p_1}}\;.
\end{equation}

To compute the average IP-WEF of the HCC ensemble, $\bar{A}_{i,p}^{\text{HCC}}$, it is possible to combine the methods that we used for PCCs and SCCs.
First, the average IP-WEF of the parallel component is computed. Then, $\bar{A}_{i,p}^{\text{HCC}}$ can be obtained by substituting $A^{\text{T}_{\text{O}}}(I,P)$ in \eqref{IPWESCC}
by the computed average IP-WEF of the parallel component  \cite{3DTurbo},

\begin{equation}
\label{IPWEHCC}
\bar{A}_{i,p}^{\text{HCC}}=\sum_{p_1}\sum_{p_2}\frac{A^{\text{T}_{\text{U}}}_{i,p_1}\cdot A^{\text{T}_{\text{L}}}_{i,p_2}\cdot A^{\text{T}_{\text{I}}}_{p_1+p_2,p-p_1-p_2}}{\binom{N}{i}\binom{2N}{p_1+p_2}}\;.
\end{equation}

It is worth mentioning that by the use of the compact graph representation, TCs can be seen as a class of protograph-based generalized LDPC (GLDPC) codes.
Therefore, equivalently, it is possible to compute the average IP-WEF of TCs by the method developed for GLDPC codes in \cite{AbuSurrahGlobeCom07,AbuSurra2011}.

\subsection{Bounds on the Error Probability}
Consider transmission of codewords of a rate-$R$ TC ensemble over the AWGN channel.
For a maximum likelihood (ML) decoder, the BER is upper bounded by

\begin{equation}
\label{BER}
P_b\leq \sum_{i=1}^{N} \sum_{p=1}^{N(1/R-1)} \frac{i}{N} \bar{A}_{i,p}\;
\mathsf{Q}\left ( \sqrt{2(i+p)R\frac{E_{\mathrm b}}{N_0}}\right).
\end{equation}
Likewise, the frame error rate (FER) is upper bounded by
\begin{equation}
\label{FER}
P_F\leq \sum_{i=1}^{N} \sum_{p=1}^{N(1/R-1)} \bar{A}_{i,p} \;\mathsf{Q}\left( \sqrt{2(i+p)R\frac{E_{\mathrm b}}{N_0}}\right),
\end{equation}
where $Q(.)$ is the $Q$-function and ${E_{\mathrm b}}/{N_0}$ is the signal-to-noise ratio.

\begin{figure}[t]
	\centering
	\definecolor{mycolor2}{rgb}{0.00000,0.54118,0.52157}%
	\definecolor{mycolor1}{rgb}{0.63529,0.07843,0.18431}%
	\begin{tikzpicture}
	
	\begin{axis}[%
	width=7cm,
	height=6.1cm,
	at={(0.794in,0.593in)},
	scale only axis,
	xmin=2,
	xmax=10,
	xlabel style={font=\color{white!15!black}},
	xlabel={$\text{E}_\text{b}\text{/N}_\text{0}$},
	ymode=log,
	ymin=1e-20,
	ymax=1,
	yminorticks=true,
	ylabel style={font=\color{white!15!black}},
	ylabel={BER},
	axis background/.style={fill=white},
	xmajorgrids,
	ymajorgrids,
	yminorgrids,
	legend style={legend cell align=left, font=\footnotesize, align=left, draw=white!15!black}
	]
	\addplot [color=mycolor1, line width=2.0pt, mark=x, mark options={solid, mycolor1}]
	table[row sep=crcr]{%
		0	1.08127674129687e+49\\
		0.5	8.1978582039498e+37\\
		1	6.70337672365652e+25\\
		1.5	6401834190216.44\\
		2	0.105701359658795\\
		2.5	5.51539025733897e-06\\
		3	2.22667984660957e-06\\
		3.5	8.52945367816259e-07\\
		4	3.04242737677132e-07\\
		4.5	9.97706591145024e-08\\
		5	2.97946570872782e-08\\
		5.5	8.0490388956193e-09\\
		6	1.96074384771405e-09\\
		6.5	4.30954488118009e-10\\
		7	8.57148726820489e-11\\
		7.5	1.54296499802588e-11\\
		8	2.48793008351179e-12\\
		8.5	3.50328114672245e-13\\
		9	4.1489315514129e-14\\
		9.5	3.95930101608258e-15\\
		10	2.91630066671554e-16\\
	};
	\addlegendentry{PCC}
	
	\addplot [color=mycolor2, line width=2.0pt, mark=triangle, mark options={solid, rotate=180, mycolor2}]
	table[row sep=crcr]{%
		0	8.0601559503983e+48\\
		0.5	5.06196825590139e+37\\
		1	2.92925123166828e+25\\
		1.5	1428167627285.4\\
		2	0.00563180017896353\\
		2.5	7.1726419084409e-10\\
		3	2.93953470547403e-10\\
		3.5	1.13727092537031e-10\\
		4	4.12601938801283e-11\\
		4.5	1.39472738324373e-11\\
		5	4.37034079667638e-12\\
		5.5	1.26699875563662e-12\\
		6	3.41275332724769e-13\\
		6.5	8.65669020861838e-14\\
		7	2.1140547829608e-14\\
		7.5	5.07123057996476e-15\\
		8	1.19447226320358e-15\\
		8.5	2.68570023713636e-16\\
		9	5.52230355494405e-17\\
		9.5	9.966986417699e-18\\
		10	1.52837105335438e-18\\
	};
	\addlegendentry{SCC}
	
	\addplot [color=red, line width=2.0pt, mark=o, mark options={solid, red}]
	table[row sep=crcr]{%
		0	1.30217371829227e+48\\
		0.5	9.08895923057792e+36\\
		1	5.90803775618208e+24\\
		1.5	328201533373.867\\
		2	0.00151217391899318\\
		2.5	1.14185277283209e-07\\
		3	8.79627275088619e-08\\
		3.5	6.59290062332492e-08\\
		4	4.79245475080335e-08\\
		4.5	3.36656880657599e-08\\
		5	2.27624082848239e-08\\
		5.5	1.47464935538338e-08\\
		6	9.10744701838694e-09\\
		6.5	5.33173476636929e-09\\
		7	2.93986148685546e-09\\
		7.5	1.51583656558528e-09\\
		8	7.25006555339804e-10\\
		8.5	3.18758111981953e-10\\
		9	1.27523986011843e-10\\
		9.5	4.58956919878121e-11\\
		10	1.46699845994459e-11\\
	};
	\addlegendentry{BCC}
	
	\addplot [color=blue, line width=2.0pt, mark size=1.7pt, mark=*, mark options={solid, blue}]
	table[row sep=crcr]{%
		0	2.91739759742634e+33\\
		0.5	1.34661005934783e+23\\
		1	422050334493.007\\
		1.5	0.0776733775763748\\
		2	6.65040171986944e-13\\
		2.5	2.89607283990355e-13\\
		3	1.19660507639199e-13\\
		3.5	4.70175065748866e-14\\
		4	1.77365416759004e-14\\
		4.5	6.52847459903677e-15\\
		5	2.39004220707178e-15\\
		5.5	8.8098829173984e-16\\
		6	3.25419898999524e-16\\
		6.5	1.17852273373363e-16\\
		7	4.06311651881909e-17\\
		7.5	1.29633042514568e-17\\
		8	3.73478307339253e-18\\
		8.5	9.50498501514067e-19\\
		9	2.09133395469223e-19\\
		9.5	3.88990075009214e-20\\
		10	5.97111341950619e-21\\
	};
	\addlegendentry{HCC}
	
	\end{axis}
	\end{tikzpicture}%
	\caption{Union bound on performance of the TCs, $K=512$, $R=1/3$.}
	\label{BERUnion}
\end{figure}
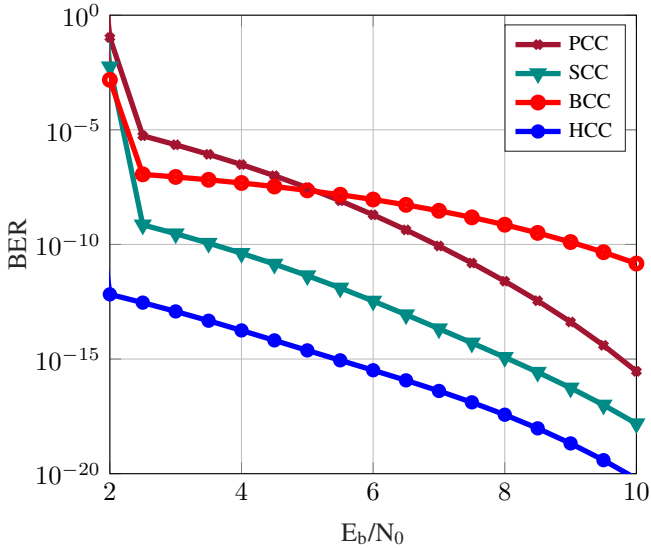
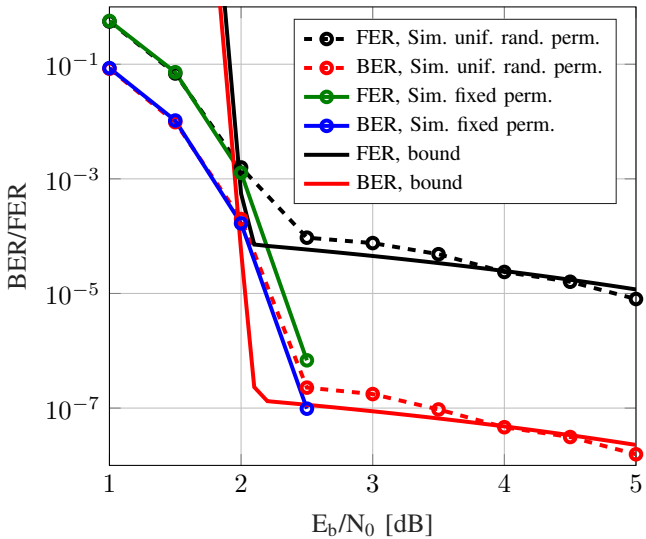
\begin{figure}[t]
	\centering
\definecolor{mycolor1}{rgb}{0.00000,0.49804,0.00000}%
\begin{tikzpicture}

\begin{axis}[%
width=7cm,
height=6.1cm,
at={(0.658in,0.589in)},
scale only axis,
xmin=1,
xmax=5,
xlabel style={font=\color{white!15!black}},
xlabel={$\text{E}_\text{b}\text{/N}_\text{0}\text{ [dB]}$},
ymode=log,
ymin=1e-08,
ymax=1,
yminorticks=true,
ylabel style={font=\color{white!15!black}},
ylabel={BER/FER},
axis background/.style={fill=white},
xmajorgrids,
ymajorgrids,
yminorgrids,
legend style={at={(0.35,0.57)}, anchor=south west, font=\footnotesize,legend cell align=left, align=left, draw=white!15!black}
]
\addplot [color=black, dashed, line width=1.5pt, mark=o, mark options={solid, black}]
table[row sep=crcr]{%
	1	0.5556\\
	1.5	0.0689\\
	2	0.00157962286504097\\
	2.5	9.41222872612207e-05\\
	3	7.57213643096815e-05\\
	3.5	4.84496124031008e-05\\
	4	2.37334261574001e-05\\
	4.5	1.60110155787182e-05\\
	5	8.00313722979408e-06\\
};
\addlegendentry{FER, Sim. unif. rand. perm.}

\addplot [color=red, dashed, line width=1.5pt, mark=o, mark options={solid, red}]
table[row sep=crcr]{%
	1	0.0834537109375\\
	1.5	0.00974296875\\
	2	0.000198918328561556\\
	2.5	2.27602257142089e-07\\
	3	1.75623281479974e-07\\
	3.5	9.46281492248062e-08\\
	4	4.6354347963672e-08\\
	4.5	3.12715148021839e-08\\
	5	1.56311274019416e-08\\
};
\addlegendentry{BER, Sim. unif. rand. perm.}

\addplot [color=mycolor1, line width=1.5pt, mark=o, mark options={solid, mycolor1}]
table[row sep=crcr]{%
	1	0.5689\\
	1.5	0.0725\\
	2	0.00127333763291495\\
	2.5	6.81625003095714e-07\\
	2.7	0\\
};
\addlegendentry{FER, Sim. fixed perm.}

\addplot [color=blue, line width=1.5pt, mark=o, mark options={solid, blue}]
table[row sep=crcr]{%
	1	0.0856236328125\\
	1.5	0.010412109375\\
	2	0.000167209540523557\\
	2.5	9.76285811725631e-08\\
	2.7	0\\
};
\addlegendentry{BER, Sim. fixed perm.}

\addplot [color=black, line width=1.5pt]
table[row sep=crcr]{%
	0	1.0008218632528e+41\\
	0.1	1.65925966128923e+39\\
	0.2	2.50582548571697e+37\\
	0.3	3.44032622418754e+35\\
	0.4	4.28531580689652e+33\\
	0.5	4.83294766727444e+31\\
	0.6	4.92481387730191e+29\\
	0.7	4.5249174147433e+27\\
	0.8	3.74078912127343e+25\\
	0.9	2.77672141109243e+23\\
	1	1.84672173021244e+21\\
	1.1	1.09813616133767e+19\\
	1.2	5.82623303214839e+16\\
	1.3	275231599560397\\
	1.4	1155328799723.91\\
	1.5	4300817273.94712\\
	1.6	14171221.2369379\\
	1.7	41256.1924203571\\
	1.8	105.941483357425\\
	1.9	0.239671293503461\\
	2	0.000550517905705948\\
	2.1	7.14961261691498e-05\\
	2.2	6.74860380463551e-05\\
	2.3	6.43902437031355e-05\\
	2.4	6.13800761618375e-05\\
	2.5	5.84546191863483e-05\\
	2.6	5.5614126069299e-05\\
	2.7	5.2858748373982e-05\\
	2.8	5.018853277967e-05\\
	2.9	4.76034199255645e-05\\
	3	4.5103243561617e-05\\
	3.1	4.26877300186215e-05\\
	3.2	4.03564980108887e-05\\
	3.3	3.81090587835473e-05\\
	3.4	3.59448166150847e-05\\
	3.5	3.38630696841501e-05\\
	3.6	3.18630113078904e-05\\
	3.7	2.99437315571898e-05\\
	3.8	2.8104219252132e-05\\
	3.9	2.63433643388166e-05\\
	4	2.46599606463495e-05\\
	4.1	2.30527090203956e-05\\
	4.2	2.15202208271634e-05\\
	4.3	2.0061021819081e-05\\
	4.4	1.86735563507652e-05\\
	4.5	1.73561919311819e-05\\
	4.6	1.610722409519e-05\\
	4.7	1.49248815749623e-05\\
	4.8	1.3807331749126e-05\\
	4.9	1.27526863448914e-05\\
	5	1.17590073659594e-05\\
	5.1	1.0824313216668e-05\\
	5.2	9.94658499066926e-06\\
	5.3	9.12377289046608e-06\\
	5.4	8.35380274240532e-06\\
	5.5	7.63458257026169e-06\\
	5.6	6.96400918937248e-06\\
	5.7	6.3399747824325e-06\\
	5.8	5.76037341755362e-06\\
	5.9	5.22310746905123e-06\\
	6	4.72609390166399e-06\\
	6.1	4.26727037954888e-06\\
	6.2	3.84460116243464e-06\\
	6.3	3.4560827527596e-06\\
	6.4	3.09974925946514e-06\\
	6.5	2.77367744635012e-06\\
	6.6	2.47599143550086e-06\\
	6.7	2.20486703927131e-06\\
	6.8	1.95853569757058e-06\\
	6.9	1.73528800078469e-06\\
	7	1.53347678247513e-06\\
	7.1	1.35151977001307e-06\\
	7.2	1.18790178547207e-06\\
	7.3	1.04117649336126e-06\\
	7.4	9.09967696074689e-07\\
	7.5	7.92970182202832e-07\\
	7.6	6.88950137036126e-07\\
	7.7	5.96745128628244e-07\\
	7.8	5.15263686618213e-07\\
	7.9	4.43484494578837e-07\\
	8	3.80455219911094e-07\\
	8.1	3.25291008192431e-07\\
	8.2	2.77172671369189e-07\\
	8.3	2.3534460122561e-07\\
	8.4	1.99112441137229e-07\\
	8.5	1.67840550207634e-07\\
	8.6	1.40949294485802e-07\\
	8.7	1.17912200067898e-07\\
	8.8	9.82530025128849e-08\\
	8.9	8.15426261656431e-08\\
	9	6.73961257130362e-08\\
	9.1	5.54696206344678e-08\\
	9.2	4.54572511927477e-08\\
	9.3	3.70881822936694e-08\\
	9.4	3.01236789781831e-08\\
	9.5	2.43542745575844e-08\\
	9.6	1.9597049519489e-08\\
	9.7	1.56930363806752e-08\\
	9.8	1.25047627010001e-08\\
	9.9	9.91394155668026e-09\\
	10	7.81931595335371e-09\\
};
\addlegendentry{FER, bound}

\addplot [color=red, line width=1.5pt]
table[row sep=crcr]{%
	0	1.73124693019867e+40\\
	0.1	2.82436588029597e+38\\
	0.2	4.1956828021137e+36\\
	0.3	5.66414377915413e+34\\
	0.4	6.93481071469053e+32\\
	0.5	7.68443689192253e+30\\
	0.6	7.69069752168299e+28\\
	0.7	6.93724536906432e+26\\
	0.8	5.62809550162749e+24\\
	0.9	4.09798716677038e+22\\
	1	2.67236018190065e+20\\
	1.1	1.55746134335849e+18\\
	1.2	8.09517185918063e+15\\
	1.3	37447321526137.8\\
	1.4	153856987861.805\\
	1.5	560344207.658673\\
	1.6	1805532.13078034\\
	1.7	5137.85905686192\\
	1.8	12.8900604152428\\
	1.9	0.0284685779150781\\
	2	5.54207410371086e-05\\
	2.1	2.32449260849562e-07\\
	2.2	1.32077340780486e-07\\
	2.3	1.25849193008051e-07\\
	2.4	1.19929697100271e-07\\
	2.5	1.14178958144694e-07\\
	2.6	1.08597197604576e-07\\
	2.7	1.03184615816798e-07\\
	2.8	9.79412052997195e-08\\
	2.9	9.28667485616159e-08\\
	3	8.79608167304609e-08\\
	3.1	8.3222768820297e-08\\
	3.2	7.86517516587704e-08\\
	3.3	7.4246700498039e-08\\
	3.4	7.00063403283588e-08\\
	3.5	6.59291879104377e-08\\
	3.6	6.20135545391339e-08\\
	3.7	5.82575495472697e-08\\
	3.8	5.4659084554252e-08\\
	3.9	5.12158784598434e-08\\
	4	4.79254631788606e-08\\
	4.1	4.47851901077857e-08\\
	4.2	4.17922373093195e-08\\
	4.3	3.89436173958056e-08\\
	4.4	3.62361860872561e-08\\
	4.5	3.3666651414457e-08\\
	4.6	3.12315835323765e-08\\
	4.7	2.89274251038905e-08\\
	4.8	2.67505022087345e-08\\
	4.9	2.46970357276568e-08\\
	5	2.27631531470211e-08\\
	5.1	2.09449007246872e-08\\
	5.2	1.92382559539098e-08\\
	5.3	1.7639140258329e-08\\
	5.4	1.6143431847929e-08\\
	5.5	1.47469786631738e-08\\
	5.6	1.34456113324526e-08\\
	5.7	1.22351560665255e-08\\
	5.8	1.11114474128994e-08\\
	5.9	1.00703407930305e-08\\
	6	9.10772474596058e-09\\
	6.1	8.21953280347929e-09\\
	6.2	7.40175492417926e-09\\
	6.3	6.65044841681976e-09\\
	6.4	5.96174828724651e-09\\
	6.5	5.33187694769191e-09\\
	6.6	4.7571532325753e-09\\
	6.7	4.23400067088467e-09\\
	6.8	3.7589549717927e-09\\
	6.9	3.32867068726759e-09\\
	7	2.93992702300214e-09\\
	7.1	2.5896327769091e-09\\
	7.2	2.27483039261599e-09\\
	7.3	1.9926991237303e-09\\
	7.4	1.7405573130212e-09\\
	7.5	1.51586379895955e-09\\
	7.6	1.31621847015486e-09\\
	7.7	1.13936199600778e-09\\
	7.8	9.83174769245422e-10\\
	7.9	8.456751028147e-10\\
	8	7.25016729775615e-10\\
	8.1	6.19485660271951e-10\\
	8.2	5.27496454282502e-10\\
	8.3	4.47587972609562e-10\\
	8.4	3.78418671394712e-10\\
	8.5	3.18761507335779e-10\\
	8.6	2.67498521700705e-10\\
	8.7	2.23615171201044e-10\\
	8.8	1.86194472824606e-10\\
	8.9	1.54411027876639e-10\\
	9	1.27524987801256e-10\\
	9.1	1.04876020925093e-10\\
	9.2	8.58773351715459e-11\\
	9.3	7.00098071364568e-11\\
	9.4	5.68162628027775e-11\\
	9.5	4.5895949716867e-11\\
	9.6	3.68992347693813e-11\\
	9.7	2.95225559360986e-11\\
	9.8	2.35036505521043e-11\\
	9.9	1.86170770244034e-11\\
	10	1.46700414326487e-11\\
};
\addlegendentry{BER, bound}

\end{axis}
\end{tikzpicture}%
	\caption{Bounds on performance of the BCCs and simulation results for uniformly random and fixed permutations, $K=512$, $R=1/3$.}
	\label{BCCFP}
\end{figure}

Fig.~\ref{BERUnion} shows the bounds on the BER performance of the different classes of TCs for $R=1/3$ and $K=512$. 
The bounds are truncated at weight $w=320$, which is larger than the corresponding Gilbert-Varshamov limit.
The HCC ensemble has the lowest error floor, while the BCC and PCC ensembles have the highest error floors.
Surprisingly, the error floor of the BCC ensemble is not only high but also has the worst slope among all TC ensembles.
On the other hand, the excellent MAP thresholds of the BCC ensemble suggest a good performance for this ensemble under MAP decoding.   
The contradiction between the excellent MAP threshold of the BCC ensemble and its poor bound suggests that the performance is dominated by few bad permutations. To verify this, we simulated the BCCs for two scenarios; first, a randomly selected but fixed set of permutations; second randomly chosen permutations for each simulated block.
The results are shown in Fig.~\ref{BCCFP}, together with the corresponding bounds. 
The figure shows that the bounds are in agreement with the simulation results for uniformly random permutations. 
However, it indicates a significant improvement in FER for the fixed set of permutations.
For example, at $E_{\mathrm b}/N_0=2.5\;\text{dB}$, the FER improves from
$9.5\cdot 10^{-5}$ to $6.8\cdot 10^{-7}$.
This significant improvement caused by fixing the permutations, supports that the high floor of the BCC ensemble is caused by the poor performance of a small fraction of codes.
Thus, in the next section, we compute expurgated union bounds.

\subsection{Bound on the Minimum Distance and Expurgated Bounds}
Consider a TC ensemble consisting of $\Omega$ codes in total. The value $\Omega$ follows from the different possible combinations of permutations and depends on the type of the ensemble. 
Assume that all codes in the ensemble are selected with equal probability. Then, the number of codewords with weight $w$ over all possible codes in the ensemble is $\Omega\bar{A}_w$, where $\bar{A}_w$ is the average WEF of the ensemble.
Therefore, given an integer value $\tilde{d}$, the total number of codewords with weight $w<\tilde{d}$ can be computed as
\[
\Omega_{w<\tilde{d}}=\Omega \sum^{\tilde{d}-1}_{w=1}\bar{A}_w\;.
\]
By considering that these codewords are spread over different possible codes, we can obtain an upper bound on the number of codes with minimum distance $d_{\text{min}}\geq\tilde{d}$,
 \begin{align}
\label{EX}
\Omega_{w\geq\tilde{d}}<\Omega-\Omega \sum^{\tilde{d}-1}_{w=1}\bar{A}_w\;.\nonumber
\end{align}
Let $\alpha$ denote the fraction of codes with $d_{\text{min}}\geq\tilde{d}$.
Then, $\alpha$ is upperbounded by
 \begin{align}
\alpha < 1-\sum^{\tilde{d}-1}_{w=1}\bar{A}_w\;.
\end{align}

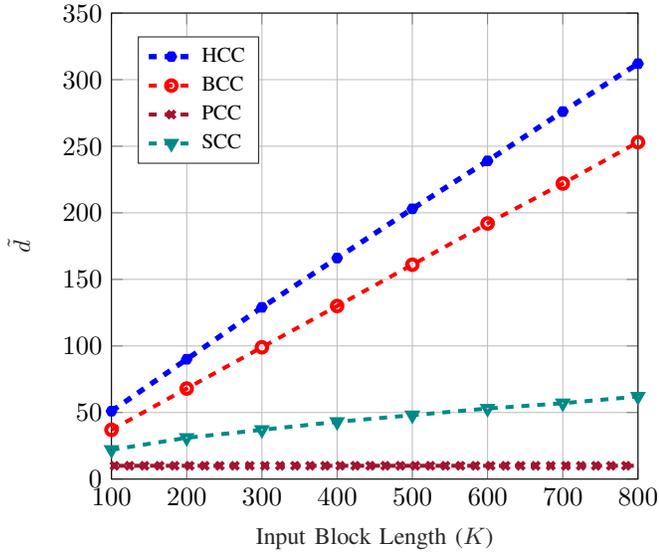
\begin{figure}[t]
	\centering
	\definecolor{mycolor1}{rgb}{0.49412,0.18431,0.55686}%
	\definecolor{mycolor2}{rgb}{0.46667,0.67451,0.18824}%
	\definecolor{mycolor3}{rgb}{0.00000,0.54118,0.52157}%
	\definecolor{mycolor4}{rgb}{0.63529,0.07843,0.18431}%
	\begin{tikzpicture}
	
	\begin{axis}[%
	width=7cm,
	height=6.2cm,
	at={(0.603in,0.492in)},
	scale only axis,
	xmin=100,
	xmax=800,
	xlabel style={font=\fontsize{9}{9}\selectfont\color{white!15!black}},
	xlabel={Input Block Length ($K$)},
	ymin=0,
	ymax=350,
	ylabel style={font=\fontsize{9}{9}\selectfont\color{white!15!black}},
	ylabel={$\tilde{d}$},
	axis background/.style={fill=white},
	xmajorgrids,
	ymajorgrids,
	legend style={at={(0.05,0.95)}, font=\footnotesize,anchor=north west, legend cell align=left, align=left, draw=white!15!black}
	]
	\addplot [color=blue, dashed, line width=2.0pt, mark=asterisk, mark options={solid, blue}]
	table[row sep=crcr]{%
		100	51\\
		200	90\\
		300	129\\
		400	166\\
		500	203\\
		600	239\\
		700	276\\
		800	312\\
		900	333\\
	};
	\addlegendentry{HCC}
	
	\addplot [color=red, dashed, line width=1.5pt, mark=o, mark options={solid, red}]
	table[row sep=crcr]{%
		100	37\\
		200	68\\
		300	99\\
		400	130\\
		500	161\\
		600	192\\
		700	222\\
		800	253\\
		900	284\\
		1000	315\\
	};
	\addlegendentry{BCC}
	
	\addplot [color=mycolor4, dashed, line width=1.5pt, mark=x, mark options={solid, mycolor4}]
	table[row sep=crcr]{%
		4	7\\
		24	9\\
		44	9\\
		64	9\\
		84	10\\
		104	10\\
		124	10\\
		144	10\\
		164	10\\
		184	10\\
		204	10\\
		224	10\\
		244	10\\
		264	10\\
		284	10\\
		304	10\\
		324	10\\
		344	10\\
		364	10\\
		384	10\\
		404	10\\
		424	10\\
		444	10\\
		464	10\\
		484	10\\
		504	10\\
		524	10\\
		544	10\\
		564	10\\
		584	10\\
		604	10\\
		624	10\\
		644	10\\
		664	10\\
		684	10\\
		704	10\\
		724	10\\
		744	10\\
		764	10\\
		784	10\\
		804	10\\
		824	10\\
		844	10\\
		864	10\\
		884	10\\
		904	10\\
		924	10\\
		944	10\\
		964	10\\
		984	10\\
	};
	\addlegendentry{PCC}
	
	\addplot [color=mycolor3, dashed, line width=1.5pt, mark=triangle, mark options={solid, rotate=180, mycolor3}]
	table[row sep=crcr]{%
		100	22\\
		200	31\\
		300	37\\
		400	43\\
		500	48\\
		600	53\\
		700	57\\
		800	62\\
		900	66\\
		1000	69\\
	};
	\addlegendentry{SCC}
	
	\end{axis}
	\end{tikzpicture}%
	\caption{Bound on the minimum distance, fraction $\alpha=0.5$ of codes have $d_{\text{min}}>\tilde{d}$.}
		\label{MinDis}
\end{figure}

For a given $\alpha$ and $\bar{A}_w$, an analytical bound on the minimum distance of an ensemble can be obtained by computing the largest $\tilde{d}$ which satisfies \eqref{EX}. 
In fact, this bound shows that a fraction $\alpha$ of all possible codes has minimum distance $d_{\text{min}}\geq\tilde{d}$.
In Fig.~\ref{MinDis}, considering different classes of TCs with $R=1/3$, this bound is computed for  $\alpha=0.5$ and different input block lengths.  
As it can be seen,  the HCC ensemble has the best minimum distance, and the PCC ensemble the worst.
As an example, for $K=300$ the values computed for HCCs, BCCs, SCCs, PCCs are 
$\tilde{d}=129$, $99$, $37$, and $10$, respectively.
Comparing the results in Fig.~\ref{MinDis} and the thresholds in Table~\ref{ThresholdsAWGN}, we can observe that the TC ensembles with good MAP threshold also have good minimum distance.
According to Fig.~\ref{MinDis}, for both the BCC and HCC ensembles, the minimum distance grows linearly with the input block length \cite{3DTurbo,MoloudiISITA}. However, the bound on the minimum distance of the HCC ensemble has a higher slope and grows faster than that of the BCC ensemble.

\begin{figure}[t]
	\centering
\definecolor{mycolor1}{rgb}{0.46667,0.67451,0.18824}%
\definecolor{mycolor2}{rgb}{0.42353,0.25098,0.39216}%
\definecolor{mycolor3}{rgb}{0.49000,0.18000,0.56000}%
\begin{tikzpicture}

\begin{axis}[%
width=7cm,
height=6.2cm,
at={(0.907in,0.664in)},
scale only axis,
xmin=0,
xmax=6,
xlabel style={font=\color{white!15!black}},
xlabel={$\text{E}_\text{b}\text{/N}_\text{0}\;\text{(dB)}$},
ymode=log,
ymin=1e-100,
ymax=1,
yminorticks=true,
ylabel style={font=\color{white!15!black}},
ylabel={Bounds on BER},
axis background/.style={fill=white},
xmajorgrids,
ymajorgrids,
yminorgrids,
legend style={at={(0.05,0.05)}, font=\footnotesize,anchor=south west, legend cell align=left, align=left, draw=white!15!black}
]
\addplot [color=blue, line width=1.5pt]
table[row sep=crcr]{%
	0	5.83479519485269e+33\\
	0.1	6.13923485406036e+31\\
	0.2	5.83908661389786e+29\\
	0.3	5.01125108754083e+27\\
	0.4	3.87404552987002e+25\\
	0.5	2.69322011869566e+23\\
	0.6	1.68101188804561e+21\\
	0.7	9.40600973748269e+18\\
	0.8	4.71162160334235e+16\\
	0.9	211019517986770\\
	1	844100668986.013\\
	1.1	3013054533.82574\\
	1.2	9591522.08511706\\
	1.3	27219.692435055\\
	1.4	68.861629285391\\
	1.5	0.155346755149862\\
	1.6	0.000312715158732166\\
	1.7	5.62303460717369e-07\\
	1.8	9.04437472937034e-10\\
	1.9	1.30354720145149e-12\\
	2	1.68690001553729e-15\\
	2.1	1.96438783046855e-18\\
	2.2	2.06324638782402e-21\\
	2.3	1.95922554143366e-24\\
	2.4	1.68590414311035e-27\\
	2.5	1.31754028706751e-30\\
	2.6	9.37128169086968e-34\\
	2.7	6.07879052104575e-37\\
	2.8	3.60302145164747e-40\\
	2.9	1.95515809403535e-43\\
	3	9.73192341646061e-47\\
	3.1	4.45217197765867e-50\\
	3.2	1.87580641831947e-53\\
	3.3	7.29417308665927e-57\\
	3.4	2.62371600784076e-60\\
	3.5	8.75084274143692e-64\\
	3.6	2.71319461946189e-67\\
	3.7	7.84342400736308e-71\\
	3.8	2.17002720632142e-74\\
	3.9	1.61330912616837e-77\\
	4	2.06553826053434e-79\\
	4.1	3.57342160245694e-81\\
	4.2	5.6663176860315e-83\\
	4.3	8.1684451078017e-85\\
	4.4	1.06805622576207e-86\\
	4.5	1.26377665436323e-88\\
	4.6	1.35005314352857e-90\\
	4.7	1.29895811051223e-92\\
	4.8	1.12289360949285e-94\\
	4.9	8.6994566733263e-97\\
	5	6.02478314057293e-99\\
	5.1	3.72002993388994e-101\\
	5.2	2.04239977055775e-103\\
	5.3	9.94330922796276e-106\\
	5.4	4.28051412043792e-108\\
	5.5	1.62474920011768e-110\\
	5.6	5.4215538371172e-113\\
	5.7	1.58562408321707e-115\\
	5.8	4.05207347843839e-118\\
	5.9	9.01956371597361e-121\\
	6	1.74310517795119e-123\\
	6.1	2.91512803716518e-126\\
	6.2	4.20455905998844e-129\\
	6.3	5.21207606470582e-132\\
	6.4	5.53340249436626e-135\\
	6.5	5.01294602962834e-138\\
	6.6	3.86104168159458e-141\\
	6.7	2.51872439912962e-144\\
	6.8	1.38623606885366e-147\\
	6.9	6.41136644825279e-151\\
	7	2.48174528965677e-154\\
	7.1	8.00667648151286e-158\\
	7.2	2.14382364570368e-161\\
	7.3	4.74326894013434e-165\\
	7.4	8.63345341852786e-169\\
	7.5	1.2868579084214e-172\\
	7.6	1.56347491423124e-176\\
	7.7	1.54096708222797e-180\\
	7.8	1.22607627463312e-184\\
	7.9	7.83598228607289e-189\\
	8	4.0022186543447e-193\\
	8.1	1.6250525887397e-197\\
	8.2	5.21756835186809e-202\\
	8.3	1.3174182697501e-206\\
	8.4	2.60135281704982e-211\\
	8.5	3.99395683639993e-216\\
	8.6	4.74008825081004e-221\\
	8.7	4.32254417001507e-226\\
	8.8	3.01018092257546e-231\\
	8.9	1.59079376083509e-236\\
	9	6.33880781040909e-242\\
	9.1	1.8919640009518e-247\\
	9.2	4.20147920523655e-253\\
	9.3	6.89413688458471e-259\\
	9.4	8.30004380157578e-265\\
	9.5	7.27891589837369e-271\\
	9.6	4.61560678144843e-277\\
	9.7	2.10030250456712e-283\\
	9.8	6.80556962841975e-290\\
	9.9	1.55789086967966e-296\\
	10	2.49907744585114e-303\\
};
\addlegendentry{HCC}

\addplot [color=red, line width=1.5pt]
table[row sep=crcr]{%
	0	5.19946636299096e+48\\
	0.1	3.67142553079613e+46\\
	0.2	2.36618501302388e+44\\
	0.3	1.39049601765046e+42\\
	0.4	7.44357856178082e+39\\
	0.5	3.62648087549128e+37\\
	0.6	1.60657109034683e+35\\
	0.7	6.46644774608445e+32\\
	0.8	2.3629069842394e+30\\
	0.9	7.83307735565289e+27\\
	1	2.35419374501037e+25\\
	1.1	6.41102600853088e+22\\
	1.2	1.58116139392232e+20\\
	1.3	3.53033856593459e+17\\
	1.4	713380528254042\\
	1.5	1304411168908.47\\
	1.6	2158122749.457\\
	1.7	3231092.64665807\\
	1.8	4378.73520130662\\
	1.9	5.37363940066829\\
	2	0.00597577489336794\\
	2.1	6.0271807583264e-06\\
	2.2	5.51994256935665e-09\\
	2.3	4.59723137014595e-12\\
	2.4	3.48816172377017e-15\\
	2.5	2.41664879907572e-18\\
	2.6	1.5329847081929e-21\\
	2.7	8.9332920089934e-25\\
	2.8	4.80150693509223e-28\\
	2.9	2.39190300749717e-31\\
	3	1.11087846209965e-34\\
	3.1	4.84471163262622e-38\\
	3.2	2.00168408171307e-41\\
	3.3	7.92249008882283e-45\\
	3.4	3.04659585672334e-48\\
	3.5	1.15989672657066e-51\\
	3.6	4.55868979992445e-55\\
	3.7	5.45037222882743e-58\\
	3.8	1.69785894590017e-59\\
	3.9	7.42682258042216e-61\\
	4	3.06104082760125e-62\\
	4.1	1.17971565451355e-63\\
	4.2	4.24049803961276e-65\\
	4.3	1.41833254373475e-66\\
	4.4	4.40446093538556e-68\\
	4.5	1.26711331428014e-69\\
	4.6	3.36986070927992e-71\\
	4.7	8.26707133872978e-73\\
	4.8	1.86681918886266e-74\\
	4.9	3.87191086647739e-76\\
	5	7.35992878865898e-78\\
	5.1	1.27934637116072e-79\\
	5.2	2.0290673181818e-81\\
	5.3	2.92961648694204e-83\\
	5.4	3.84172723218442e-85\\
	5.5	4.5647929663537e-87\\
	5.6	4.90289478870584e-89\\
	5.7	4.74852597356404e-91\\
	5.8	4.13672221225056e-93\\
	5.9	3.23325178134048e-95\\
	6	2.26141129962159e-97\\
	6.1	1.41163934124018e-99\\
	6.2	7.84322448602029e-102\\
	6.3	3.86802433103734e-104\\
	6.4	1.68841035307654e-106\\
	6.5	6.50435921498645e-109\\
	6.6	2.20487634976498e-111\\
	6.7	6.55698146648514e-114\\
	6.8	1.70537914185e-116\\
	6.9	3.86689330689564e-119\\
	7	7.61944894113244e-122\\
	7.1	1.30037628848041e-124\\
	7.2	1.91571048562996e-127\\
	7.3	2.42774890030789e-130\\
	7.4	2.6372664956819e-133\\
	7.5	2.44685756451579e-136\\
	7.6	1.93178914822759e-139\\
	7.7	1.29289142348077e-142\\
	7.8	7.30689252900021e-146\\
	7.9	3.47336102735936e-149\\
	8	1.38309536042692e-152\\
	8.1	4.59448716831765e-156\\
	8.2	1.26782911289723e-159\\
	8.3	2.89358673667953e-163\\
	8.4	5.43793536159262e-167\\
	8.5	8.376801555602e-171\\
	8.6	1.05280553233334e-174\\
	8.7	1.07442484188567e-178\\
	8.8	8.86025225829176e-183\\
	8.9	5.87480397802201e-187\\
	9	3.11603784873149e-191\\
	9.1	1.3152474147846e-195\\
	9.2	4.39428524632144e-200\\
	9.3	1.15577114694776e-204\\
	9.4	2.37974190945539e-209\\
	9.5	3.81397120066802e-214\\
	9.6	4.73011668215266e-219\\
	9.7	4.51243979339962e-224\\
	9.8	3.29104506936198e-229\\
	9.9	1.82353828844933e-234\\
	10	7.62723339431275e-240\\
};
\addlegendentry{BCC}

\addplot [color=mycolor1, line width=1.5pt]
table[row sep=crcr]{%
	0	1.6120311857813e+49\\
	0.1	1.11548338086192e+47\\
	0.2	7.04257642006771e+44\\
	0.3	4.0527091666685e+42\\
	0.4	2.12366348694636e+40\\
	0.5	1.01239364873186e+38\\
	0.6	4.38683747271313e+35\\
	0.7	1.72634251502533e+33\\
	0.8	6.16499310393967e+30\\
	0.9	1.99640999745881e+28\\
	1	5.85850245102297e+25\\
	1.1	1.55697322667991e+23\\
	1.2	3.74544928782908e+20\\
	1.3	8.15195017551442e+17\\
	1.4	1.60472294250766e+15\\
	1.5	2856335249695.57\\
	1.6	4596489810.9199\\
	1.7	6687155.46391923\\
	1.8	8796.34645077746\\
	1.9	10.4645464982419\\
	2	0.0112635970182631\\
	2.1	1.0975669690812e-05\\
	2.2	9.69005034569941e-09\\
	2.3	7.7590301870022e-12\\
	2.4	5.70764182303822e-15\\
	2.5	3.67032505500189e-17\\
	2.6	1.61974133500832e-17\\
	2.7	7.83270119462798e-18\\
	2.8	3.72881783978889e-18\\
	2.9	1.74662266351181e-18\\
	3	8.04685739933093e-19\\
	3.1	3.64485398127067e-19\\
	3.2	1.62249767057576e-19\\
	3.3	7.09508700058834e-20\\
	3.4	3.0466082351264e-20\\
	3.5	1.28402107704624e-20\\
	3.6	5.30922970366664e-21\\
	3.7	2.15276941115872e-21\\
	3.8	8.55595510211202e-22\\
	3.9	3.33148786503905e-22\\
	4	1.27026931725775e-22\\
	4.1	4.74051452056113e-23\\
	4.2	1.73063942120689e-23\\
	4.3	6.17750954481503e-24\\
	4.4	2.15484029495909e-24\\
	4.5	7.34134079502365e-25\\
	4.6	2.44147727542226e-25\\
	4.7	7.92136083078206e-26\\
	4.8	2.50590047975609e-26\\
	4.9	7.72477820848361e-27\\
	5	2.31900497211866e-27\\
	5.1	6.77547834646448e-28\\
	5.2	1.92541234778057e-28\\
	5.3	5.31825942503594e-29\\
	5.4	1.42687900679609e-29\\
	5.5	3.71603501925323e-30\\
	5.6	9.38736551519662e-31\\
	5.7	2.29862507786088e-31\\
	5.8	5.45173840384027e-32\\
	5.9	1.25146557034817e-32\\
	6	2.7783492091298e-33\\
	6.1	5.96071044314405e-34\\
	6.2	1.23482038513127e-34\\
	6.3	2.46801218205785e-35\\
	6.4	4.75514103362221e-36\\
	6.5	8.8242627808218e-37\\
	6.6	1.57583231452781e-37\\
	6.7	2.70561996856849e-38\\
	6.8	4.46219310000979e-39\\
	6.9	7.06227930737044e-40\\
	7	1.07161060193487e-40\\
	7.1	1.55738330235159e-41\\
	7.2	2.16561480766782e-42\\
	7.3	2.87835670403569e-43\\
	7.4	3.65279856920901e-44\\
	7.5	4.42133877445022e-45\\
	7.6	5.09855440845122e-46\\
	7.7	5.59515971425709e-47\\
	7.8	5.83640628426524e-48\\
	7.9	5.78002573686136e-49\\
	8	5.42796204728102e-50\\
	8.1	4.82753897593406e-51\\
	8.2	4.06109602900424e-52\\
	8.3	3.22717703107475e-53\\
	8.4	2.41927203530879e-54\\
	8.5	1.70858176532004e-55\\
	8.6	1.13519078661361e-56\\
	8.7	7.0853720172004e-58\\
	8.8	4.14840659056605e-59\\
	8.9	2.27496179953589e-60\\
	9	1.16674271649233e-61\\
	9.1	5.5873101329023e-63\\
	9.2	2.49436331868626e-64\\
	9.3	1.03641135484552e-65\\
	9.4	4.00119503540288e-67\\
	9.5	1.43279783561412e-68\\
	9.6	4.75064768477511e-70\\
	9.7	1.45582998391759e-71\\
	9.8	4.11582949094486e-73\\
	9.9	1.07145325690782e-74\\
	10	2.56341762458841e-76\\
};
\addlegendentry{SCC}

\addplot [color=mycolor2, line width=1.5pt]
table[row sep=crcr]{%
	0	2.16255348259374e+49\\
	0.1	1.54123251200281e+47\\
	0.2	1.00575121946955e+45\\
	0.3	6.00639695678386e+42\\
	0.4	3.2814535551403e+40\\
	0.5	1.63957164078996e+38\\
	0.6	7.49124618785127e+35\\
	0.7	3.13011427831488e+33\\
	0.8	1.19635460117273e+31\\
	0.9	4.18473863917531e+28\\
	1	1.3406753447313e+26\\
	1.1	3.93831815717488e+23\\
	1.2	1.06240829985288e+21\\
	1.3	2.63719077753425e+18\\
	1.4	6.03948987997895e+15\\
	1.5	12803668380432.9\\
	1.6	25236288225.2094\\
	1.7	46503996.2581599\\
	1.8	80696.5013460045\\
	1.9	133.111365237132\\
	2	0.211401432601363\\
	2.1	0.000350126546946405\\
	2.2	1.82693673475069e-05\\
	2.3	1.48167486940589e-05\\
	2.4	1.23636711307378e-05\\
	2.5	1.03119420204021e-05\\
	2.6	8.59241377929651e-06\\
	2.7	7.15011292634068e-06\\
	2.8	5.94017334581975e-06\\
	2.9	4.92557922569737e-06\\
	3	4.0755351188306e-06\\
	3.1	3.36425498133596e-06\\
	3.2	2.77004130274897e-06\\
	3.3	2.27457063376989e-06\\
	3.4	1.86232942201686e-06\\
	3.5	1.52016158051501e-06\\
	3.6	1.23690067107028e-06\\
	3.7	1.00306727251744e-06\\
	3.8	8.10617373980805e-07\\
	3.9	6.52731312714899e-07\\
	4	5.23635385733288e-07\\
	4.1	4.18450140923433e-07\\
	4.2	3.33060719159062e-07\\
	4.3	2.6400562404114e-07\\
	4.4	2.08381043224987e-07\\
	4.5	1.63758406524887e-07\\
	4.6	1.28113291764047e-07\\
	4.7	9.97641158286835e-08\\
	4.8	7.73193016756013e-08\\
	4.9	5.96318110872112e-08\\
	5	4.57600916168454e-08\\
	5.1	3.49346145796602e-08\\
	5.2	2.65292865868796e-08\\
	5.3	2.0037105483476e-08\\
	5.4	1.50495066634707e-08\\
	5.5	1.12389105575664e-08\\
	5.6	8.34403877352645e-09\\
	5.7	6.15761649297486e-09\\
	5.8	4.51612343188422e-09\\
	5.9	3.2912957866349e-09\\
	6	2.38311771987803e-09\\
	6.1	1.71407353904309e-09\\
	6.2	1.22446112438068e-09\\
	6.3	8.68593638591565e-10\\
	6.4	6.11740314428723e-10\\
	6.5	4.27678358896191e-10\\
	6.6	2.96746904676696e-10\\
	6.7	2.04310644294952e-10\\
	6.8	1.39555459406175e-10\\
	6.9	9.45511640994576e-11\\
	7	6.35275740357598e-11\\
	7.1	4.23196482037617e-11\\
	7.2	2.79455803458921e-11\\
	7.3	1.82885921215516e-11\\
	7.4	1.18589438175795e-11\\
	7.5	7.61746784919928e-12\\
	7.6	4.84587398687389e-12\\
	7.7	3.05229227366276e-12\\
	7.8	1.90311877182456e-12\\
	7.9	1.17430868682089e-12\\
	8	7.16908306724237e-13\\
	8.1	4.3290812868179e-13\\
	8.2	2.58500815456958e-13\\
	8.3	1.52595871105821e-13\\
	8.4	8.90258862909428e-14\\
	8.5	5.13165834434265e-14\\
	8.6	2.92172636151412e-14\\
	8.7	1.64259827020403e-14\\
	8.8	9.11590841600787e-15\\
	8.9	4.99239709134236e-15\\
	9	2.6972404454403e-15\\
	9.1	1.43710766316505e-15\\
	9.2	7.54870817106147e-16\\
	9.3	3.9076984240923e-16\\
	9.4	1.99288676827189e-16\\
	9.5	1.00092640438462e-16\\
	9.6	4.94903972208744e-17\\
	9.7	2.40811283583621e-17\\
	9.8	1.15266755173696e-17\\
	9.9	5.42541816357987e-18\\
	10	2.51010041225898e-18\\
};
\addlegendentry{PCC}

\addplot [color=blue, dashed, line width=1.5pt, forget plot]
table[row sep=crcr]{%
	0	1.18950384324866e-33\\
	0.1	2.30615449671337e-34\\
	0.2	4.30705375937924e-35\\
	0.3	7.74234047735244e-36\\
	0.4	1.3384017461128e-36\\
	0.5	2.22298416442991e-37\\
	0.6	3.5442656224782e-38\\
	0.7	5.41940103864e-39\\
	0.8	7.93957149787726e-40\\
	0.9	1.11336979885566e-40\\
	1	1.49294648667587e-41\\
	1.1	1.91234842981144e-42\\
	1.2	2.33750617783384e-43\\
	1.3	2.72355162666594e-44\\
	1.4	3.02161108905061e-45\\
	1.5	3.18840301272468e-46\\
	1.6	3.196240166682e-47\\
	1.7	3.04035975206902e-48\\
	1.8	2.74097709301399e-49\\
	1.9	2.33907166404341e-50\\
	2	1.88707513534634e-51\\
	2.1	1.43740433385421e-52\\
	2.2	1.03237082091195e-53\\
	2.3	6.98181702219985e-55\\
	2.4	4.43988613304187e-56\\
	2.5	2.65109839298631e-57\\
	2.6	1.48421063213323e-58\\
	2.7	7.77912246215601e-60\\
	2.8	3.81122925340261e-61\\
	2.9	1.74267704212389e-62\\
	3	7.42485018423727e-64\\
	3.1	2.94281011932431e-65\\
	3.2	1.08319893418848e-66\\
	3.3	3.69637318649993e-68\\
	3.4	1.16733975555512e-69\\
	3.5	3.40554491528507e-71\\
	3.6	9.16092095886201e-73\\
	3.7	2.26793750286297e-74\\
	3.8	5.15726199733905e-76\\
	3.9	1.07507880609578e-77\\
	4	2.05026916426625e-79\\
	4.1	3.56965813270855e-81\\
	4.2	5.66189727828599e-83\\
	4.3	8.1633741381744e-85\\
	4.4	1.06753058311844e-86\\
	4.5	1.26328615367619e-88\\
	4.6	1.3496421318047e-90\\
	4.7	1.29864962421431e-92\\
	4.8	1.12268675677297e-94\\
	4.9	8.69822077946773e-97\\
	5	6.02412696474684e-99\\
	5.1	3.71972120596321e-101\\
	5.2	2.04227141399023e-103\\
	5.3	9.94283902185298e-106\\
	5.4	4.28036279953993e-108\\
	5.5	1.62470654876967e-110\\
	5.6	5.42144887269553e-113\\
	5.7	1.58560160053634e-115\\
	5.8	4.05203170118498e-118\\
	5.9	9.01949659256492e-121\\
	6	1.74309588457536e-123\\
	6.1	2.91511698805826e-126\\
	6.2	4.20454781940739e-129\\
	6.3	5.21206631534615e-132\\
	6.4	5.53339531195507e-135\\
	6.5	5.01294155234315e-138\\
	6.6	3.86103932917621e-141\\
	6.7	2.51872336151698e-144\\
	6.8	1.38623568620333e-147\\
	6.9	6.41136527335782e-151\\
	7	2.48174499059022e-154\\
	7.1	8.00667585315514e-158\\
	7.2	2.14382353721844e-161\\
	7.3	4.74326878693114e-165\\
	7.4	8.63345324238662e-169\\
	7.5	1.28685789201302e-172\\
	7.6	1.56347490190735e-176\\
	7.7	1.54096707480248e-180\\
	7.8	1.22607627106234e-184\\
	7.9	7.83598227244038e-189\\
	8	4.00221865023487e-193\\
	8.1	1.6250525877667e-197\\
	8.2	5.21756835006923e-202\\
	8.3	1.3174182694919e-206\\
	8.4	2.60135281676376e-211\\
	8.5	3.9939568361568e-216\\
	8.6	4.74008825065249e-221\\
	8.7	4.32254416993771e-226\\
	8.8	3.01018092254687e-231\\
	8.9	1.59079376082719e-236\\
	9	6.33880781039288e-242\\
	9.1	1.89196400094934e-247\\
	9.2	4.20147920523383e-253\\
	9.3	6.89413688458252e-259\\
	9.4	8.30004380157451e-265\\
	9.5	7.27891589837317e-271\\
	9.6	4.61560678144828e-277\\
	9.7	2.10030250456709e-283\\
	9.8	6.80556962841971e-290\\
	9.9	1.55789086967965e-296\\
	10	2.49907744585113e-303\\
};
\addplot [color=red, dashed, line width=1.5pt, forget plot]
table[row sep=crcr]{%
	0	2.57541578402475e-25\\
	0.1	6.79265259549529e-26\\
	0.2	1.7380449116319e-26\\
	0.3	4.31140545379372e-27\\
	0.4	1.03612901288272e-27\\
	0.5	2.41068629857872e-28\\
	0.6	5.42610458656377e-29\\
	0.7	1.18069771446604e-29\\
	0.8	2.48181041459725e-30\\
	0.9	5.03555831557572e-31\\
	1	9.85460316550156e-32\\
	1.1	1.85865963276373e-32\\
	1.2	3.37582415950979e-33\\
	1.3	5.89959424132829e-34\\
	1.4	9.91203891092346e-35\\
	1.5	1.59967360066548e-35\\
	1.6	2.47770126135677e-36\\
	1.7	3.6798474852713e-37\\
	1.8	5.23576788880965e-38\\
	1.9	7.13015570782312e-39\\
	2	9.28486943518019e-40\\
	2.1	1.15502730534722e-40\\
	2.2	1.37126355156268e-41\\
	2.3	1.55211926607345e-42\\
	2.4	1.67323992589317e-43\\
	2.5	1.7161830166145e-44\\
	2.6	1.67291466004298e-45\\
	2.7	1.54813938514716e-46\\
	2.8	1.35856941285645e-47\\
	2.9	1.12923943314722e-48\\
	3	8.87986691741987e-50\\
	3.1	6.59801155667545e-51\\
	3.2	4.62659355272535e-52\\
	3.3	3.05767662065644e-53\\
	3.4	1.90207683622735e-54\\
	3.5	1.1121891669453e-55\\
	3.6	6.10427287951445e-57\\
	3.7	3.14026737035771e-58\\
	3.8	1.51192774481885e-59\\
	3.9	6.80240590880004e-61\\
	4	2.85546675627301e-62\\
	4.1	1.1165312188412e-63\\
	4.2	4.05994164515865e-65\\
	4.3	1.3705009944069e-66\\
	4.4	4.28730440745706e-68\\
	4.5	1.24064781456779e-69\\
	4.6	3.31485596585502e-71\\
	4.7	8.16214372958866e-73\\
	4.8	1.84849097318217e-74\\
	4.9	3.84266507125983e-76\\
	5	7.31739958716991e-78\\
	5.1	1.27372359292675e-79\\
	5.2	2.02232510359566e-81\\
	5.3	2.92230213088542e-83\\
	5.4	3.8345659704843e-85\\
	5.5	4.55848131288582e-87\\
	5.6	4.89789999184536e-89\\
	5.7	4.74498621323765e-91\\
	5.8	4.13448168826469e-93\\
	5.9	3.23198861542529e-95\\
	6	2.26077874572969e-97\\
	6.1	1.41135878030503e-99\\
	6.2	7.84212550654947e-102\\
	6.3	3.86764527981583e-104\\
	6.4	1.68829558065408e-106\\
	6.5	6.50405508084114e-109\\
	6.6	2.20480604126066e-111\\
	6.7	6.55684012743264e-114\\
	6.8	1.70535451587728e-116\\
	6.9	3.86685624440664e-119\\
	7	7.61940092515879e-122\\
	7.1	1.30037095251599e-124\\
	7.2	1.91570541750439e-127\\
	7.3	2.42774480124513e-130\\
	7.4	2.63726368324103e-133\\
	7.5	2.44685593384916e-136\\
	7.6	1.93178835241146e-139\\
	7.7	1.29289109789196e-142\\
	7.8	7.30689141691432e-146\\
	7.9	3.47336071158206e-149\\
	8	1.3830952862077e-152\\
	8.1	4.59448702456344e-156\\
	8.2	1.26782909005573e-159\\
	8.3	2.89358670704389e-163\\
	8.4	5.43793533034413e-167\\
	8.5	8.3768015289543e-171\\
	8.6	1.05280553050461e-174\\
	8.7	1.07442484088085e-178\\
	8.8	8.86025225389409e-183\\
	8.9	5.87480397649712e-187\\
	9	3.11603784831485e-191\\
	9.1	1.31524741469539e-195\\
	9.2	4.39428524617263e-200\\
	9.3	1.15577114692853e-204\\
	9.4	2.37974190943626e-209\\
	9.5	3.81397120065345e-214\\
	9.6	4.73011668214423e-219\\
	9.7	4.51243979339593e-224\\
	9.8	3.29104506936077e-229\\
	9.9	1.82353828844903e-234\\
	10	7.62723339431222e-240\\
};
\addplot [color=mycolor1, dashed, line width=1.5pt, forget plot]
table[row sep=crcr]{%
	0	2.6009146182617e-11\\
	0.1	1.71657865720567e-11\\
	0.2	1.12283466105641e-11\\
	0.3	7.27776715942125e-12\\
	0.4	4.67330912520809e-12\\
	0.5	2.97239533961525e-12\\
	0.6	1.87220726257135e-12\\
	0.7	1.16754589604133e-12\\
	0.8	7.20728690053096e-13\\
	0.9	4.40301607511207e-13\\
	1	2.66139746217674e-13\\
	1.1	1.59128558357685e-13\\
	1.2	9.40937552329903e-14\\
	1.3	5.50098532225362e-14\\
	1.4	3.17890244596961e-14\\
	1.5	1.81534081462463e-14\\
	1.6	1.02415987844464e-14\\
	1.7	5.70673052567462e-15\\
	1.8	3.13975215559137e-15\\
	1.9	1.70516740848603e-15\\
	2	9.13848906400505e-16\\
	2.1	4.83154770211427e-16\\
	2.2	2.51923147129994e-16\\
	2.3	1.29503582231687e-16\\
	2.4	6.56123894332708e-17\\
	2.5	3.27517974012821e-17\\
	2.6	1.61020361179059e-17\\
	2.7	7.7941761310377e-18\\
	2.8	3.71318794648862e-18\\
	2.9	1.74040589448466e-18\\
	3	8.02262768757516e-19\\
	3.1	3.63560498445324e-19\\
	3.2	1.61904156251609e-19\\
	3.3	7.08245113651852e-20\\
	3.4	3.04209049483302e-20\\
	3.5	1.2824423576423e-20\\
	3.6	5.30384056318301e-21\\
	3.7	2.1509733417317e-21\\
	3.8	8.5501143534078e-22\\
	3.9	3.32963560123855e-22\\
	4	1.26969682783662e-22\\
	4.1	4.73879107365602e-23\\
	4.2	1.73013438368904e-23\\
	4.3	6.17606985822953e-24\\
	4.4	2.15444131800276e-24\\
	4.5	7.3402666234146e-25\\
	4.6	2.44119650630341e-25\\
	4.7	7.92064884418635e-26\\
	4.8	2.50572544224891e-26\\
	4.9	7.72436132939586e-27\\
	5	2.31890885863294e-27\\
	5.1	6.775263997291e-28\\
	5.2	1.92536614342443e-28\\
	5.3	5.31816323720156e-29\\
	5.4	1.42685968364627e-29\\
	5.5	3.7159975916142e-30\\
	5.6	9.38729567725652e-31\\
	5.7	2.29861253507576e-31\\
	5.8	5.4517167412515e-32\\
	5.9	1.25146197582143e-32\\
	6	2.7783434840708e-33\\
	6.1	5.96070169919743e-34\\
	6.2	1.23481910575886e-34\\
	6.3	2.46801039058125e-35\\
	6.4	4.75513863532781e-36\\
	6.5	8.82425971453245e-37\\
	6.6	1.57583194052534e-37\\
	6.7	2.7056195338511e-38\\
	6.8	4.46219261903754e-39\\
	6.9	7.06227880141606e-40\\
	7	1.07161055139115e-40\\
	7.1	1.55738325445975e-41\\
	7.2	2.16561476467892e-42\\
	7.3	2.87835666752662e-43\\
	7.4	3.65279853991145e-44\\
	7.5	4.42133875226461e-45\\
	7.6	5.09855439261941e-46\\
	7.7	5.59515970362534e-47\\
	7.8	5.83640627755596e-48\\
	7.9	5.78002573288842e-49\\
	8	5.42796204507675e-50\\
	8.1	4.82753897478994e-51\\
	8.2	4.06109602844954e-52\\
	8.3	3.22717703082395e-53\\
	8.4	2.41927203520321e-54\\
	8.5	1.70858176527873e-55\\
	8.6	1.13519078659862e-56\\
	8.7	7.08537201714997e-58\\
	8.8	4.14840659055037e-59\\
	8.9	2.27496179953139e-60\\
	9	1.16674271649114e-61\\
	9.1	5.58731013289941e-63\\
	9.2	2.49436331868562e-64\\
	9.3	1.03641135484539e-65\\
	9.4	4.00119503540264e-67\\
	9.5	1.43279783561408e-68\\
	9.6	4.75064768477505e-70\\
	9.7	1.45582998391758e-71\\
	9.8	4.11582949094485e-73\\
	9.9	1.07145325690782e-74\\
	10	2.56341762458841e-76\\
};
\addplot [color=mycolor2, dashed, line width=1.5pt, forget plot]
table[row sep=crcr]{%
	0	0.000348335795198516\\
	0.1	0.000309520520929779\\
	0.2	0.000274517458942697\\
	0.3	0.000243013837331099\\
	0.4	0.000214715623156207\\
	0.5	0.00018934712811598\\
	0.6	0.000166650544830509\\
	0.7	0.000146385421580615\\
	0.8	0.00012832808348894\\
	0.9	0.000112271008178121\\
	1	9.80221638826874e-05\\
	1.1	8.54043178358623e-05\\
	1.2	7.42543225064377e-05\\
	1.3	6.44223869331697e-05\\
	1.4	5.57713400043822e-05\\
	1.5	4.81758920695288e-05\\
	1.6	4.15219007587266e-05\\
	1.7	3.57056463375095e-05\\
	1.8	3.06331213490978e-05\\
	1.9	2.62193387069086e-05\\
	2	2.23876618069726e-05\\
	2.1	1.90691596436986e-05\\
	2.2	1.6201989342471e-05\\
	2.3	1.3730807977158e-05\\
	2.4	1.16062150267924e-05\\
	2.5	9.78422634919442e-06\\
	2.6	8.22578011448883e-06\\
	2.7	6.89627475134197e-06\\
	2.8	5.76513861501098e-06\\
	2.9	4.80543078945266e-06\\
	3	3.99347218516878e-06\\
	3.1	3.30850588859996e-06\\
	3.2	2.73238555534659e-06\\
	3.3	2.24929051527679e-06\\
	3.4	1.84546616919541e-06\\
	3.5	1.50898820039715e-06\\
	3.6	1.22954909616284e-06\\
	3.7	9.9826547010733e-07\\
	3.8	8.07504692380689e-07\\
	3.9	6.50729367384871e-07\\
	4	5.22358244489993e-07\\
	4.1	4.17642203139109e-07\\
	4.2	3.32554016991878e-07\\
	4.3	2.63690670021494e-07\\
	4.4	2.0818706875647e-07\\
	4.5	1.6364006750937e-07\\
	4.6	1.28041796141075e-07\\
	4.7	9.97213516427582e-08\\
	4.8	7.72939847998844e-08\\
	4.9	5.9616980866161e-08\\
	5	4.57514981259592e-08\\
	5.1	3.49296902036715e-08\\
	5.2	2.65264968350277e-08\\
	5.3	2.00355434732278e-08\\
	5.4	1.5048642541416e-08\\
	5.5	1.12384383794265e-08\\
	5.6	8.34378400809923e-09\\
	5.7	6.15748080522081e-09\\
	5.8	4.51605211942056e-09\\
	5.9	3.29125881478495e-09\\
	6	2.38309881772143e-09\\
	6.1	1.71406401247467e-09\\
	6.2	1.22445639293286e-09\\
	6.3	8.68591323712195e-10\\
	6.4	6.11739199157539e-10\\
	6.5	4.2767782997848e-10\\
	6.6	2.96746657853767e-10\\
	6.7	2.04310531001882e-10\\
	6.8	1.39555408277013e-10\\
	6.9	9.45511414214572e-11\\
	7	6.35275641540752e-11\\
	7.1	4.23196439754911e-11\\
	7.2	2.7945578570021e-11\\
	7.3	1.82885913897655e-11\\
	7.4	1.18589435218575e-11\\
	7.5	7.61746773205847e-12\\
	7.6	4.84587394141155e-12\\
	7.7	3.05229225638436e-12\\
	7.8	1.90311876539698e-12\\
	7.9	1.17430868448172e-12\\
	8	7.16908305891851e-13\\
	8.1	4.32908128392321e-13\\
	8.2	2.58500815358635e-13\\
	8.3	1.52595871073219e-13\\
	8.4	8.90258862803958e-14\\
	8.5	5.13165834400995e-14\\
	8.6	2.92172636141185e-14\\
	8.7	1.64259827017341e-14\\
	8.8	9.11590841591866e-15\\
	8.9	4.99239709131708e-15\\
	9	2.69724044543334e-15\\
	9.1	1.43710766316318e-15\\
	9.2	7.54870817105663e-16\\
	9.3	3.90769842409109e-16\\
	9.4	1.9928867682716e-16\\
	9.5	1.00092640438455e-16\\
	9.6	4.94903972208727e-17\\
	9.7	2.40811283583617e-17\\
	9.8	1.15266755173695e-17\\
	9.9	5.42541816357986e-18\\
	10	2.51010041225898e-18\\
};
\addplot [color=blue, dotted, line width=1.5pt, mark=asterisk, mark options={solid, blue}, forget plot]
table[row sep=crcr]{%
	3	1.19660507639199e-13\\
	3.5	4.70175065748866e-14\\
	4	1.77365416759004e-14\\
	4.5	6.52847459903677e-15\\
	5	2.39004220707178e-15\\
	5.5	8.8098829173984e-16\\
	6	3.25419898999524e-16\\
	6.5	1.17852273373363e-16\\
	7	4.06311651881909e-17\\
	7.5	1.29633042514568e-17\\
	8	3.73478307339253e-18\\
	8.5	9.50498501514067e-19\\
	9	2.09133395469223e-19\\
	9.5	3.88990075009214e-20\\
	10	5.97111341950619e-21\\
};
\addplot [color=mycolor1, dotted, line width=1.5pt, mark=triangle, mark options={solid, mycolor1}, forget plot]
table[row sep=crcr]{%
	3	2.93953470547403e-10\\
	3.5	1.13727092537031e-10\\
	4	4.12601938801283e-11\\
	4.5	1.39472738324373e-11\\
	5	4.37034079667638e-12\\
	5.5	1.26699875563662e-12\\
	6	3.41275332724769e-13\\
	6.5	8.65669020861838e-14\\
	7	2.1140547829608e-14\\
	7.5	5.07123057996476e-15\\
	8	1.19447226320358e-15\\
	8.5	2.68570023713636e-16\\
	9	5.52230355494405e-17\\
	9.5	9.966986417699e-18\\
	10	1.52837105335438e-18\\
};
\addplot [color=red, dotted, line width=1.5pt, mark=o, mark options={solid, red}, forget plot]
table[row sep=crcr]{%
	3	9.00738329690746e-08\\
	3.5	6.75113023828472e-08\\
	4	4.90747366482262e-08\\
	4.5	3.44736645793382e-08\\
	5	2.33087060836597e-08\\
	5.5	1.51004093991258e-08\\
	6	9.32602574682822e-09\\
	6.5	5.45969640076215e-09\\
	7	3.01041816253999e-09\\
	7.5	1.55221664315933e-09\\
	8	7.42406712667959e-10\\
	8.5	3.26408306669519e-10\\
	9	1.30584561676127e-10\\
	9.5	4.69971885955196e-11\\
	10	1.50220642298326e-11\\
};
\addplot [color=mycolor3, dotted, line width=1.5pt, mark=x, mark options={solid, mycolor3}, forget plot]
table[row sep=crcr]{%
	3	2.28997965256643e-06\\
	3.5	8.7723490805736e-07\\
	4	3.13022560386828e-07\\
	4.5	1.02719015394744e-07\\
	5	3.07065281936111e-08\\
	5.5	8.30744199352229e-09\\
	6	2.02766522230175e-09\\
	6.5	4.46769952987495e-10\\
	7	8.91130630934014e-11\\
	7.5	1.60867171341276e-11\\
	8	2.60000773307729e-12\\
	8.5	3.66708746363477e-13\\
	9	4.34702279292935e-14\\
	9.5	4.1502466372239e-15\\
	10	3.05753288238676e-16\\
};
\end{axis}

\begin{axis}[%
width=7cm,
height=6.2cm,
at={(0.907in,0.664in)},
scale only axis,
xmin=0,
xmax=6,
ymin=0,
ymax=6,
axis line style={draw=none},
ticks=none,
axis x line*=bottom,
axis y line*=left
]
\draw [black, thick] (axis cs:1,4.6) ellipse [x radius=10, y radius=140];
\draw [black,thick] (axis cs:5,5.4) ellipse [x radius=10, y radius=50];
\node[below right, align=left, draw=white]
at (rel axis cs:0.01,0.5) {Expurgated Bounds};
\node[below right, align=left]
at (rel axis cs:0.615,0.83) {Union Bounds};
\end{axis}
\end{tikzpicture}%
	\caption{Expurgated union bound on performance of TCs, $\alpha=0.5$, $K=512$, and $R=1/3$.}
	\label{EXBER}
\end{figure}
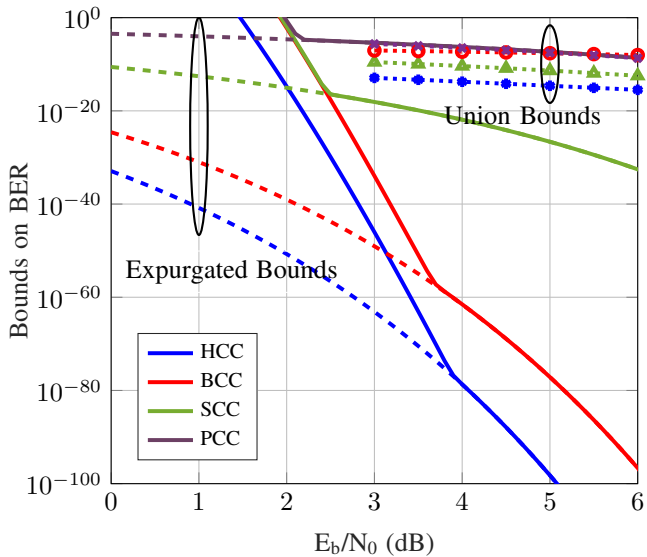

Consider excluding a fraction $(1-\alpha)$ of codes with $d_{\text{min}}<\tilde{d}$ from a TC ensemble. Then, it is possible to compute the upper bound on the performance of this expurgated ensemble. The average BER of the expurgated ensemble is upperbounded by
\begin{equation}
P_b\leq \frac{1}{\alpha}\mathop{\sum_{i=1}^{kN}\; \sum_{w=\tilde{d}}^{nN}}\;\frac{i}{N} \bar{A}_{i,w}\;
\mathsf{Q}\left ( \sqrt{2wR\frac{E_{\mathrm b}}{N_0}}\right).
\end{equation}
The bounds for the expurgated TC ensembles are shown in Fig.~\ref{EXBER} for $\alpha=0.5$, which means that half of the codes with $d_{\text{min}}<\tilde{d}(\alpha)$ are excluded. For comparison, we also provide the corresponding union bounds in the same figure.
It can be seen that for all TC ensembles except the PCC ensemble, the error floor estimated by the expurgated bound is much steeper and lower than that resulting from the union bound.
In other words, expurgation improves the performance of the SCC, BCC, and HCC ensembles significantly.

\begin{figure}[t]
	\centering
\begin{tikzpicture}

\begin{axis}[%
width=7cm,
height=6.2cm,
at={(0.788in,0.595in)},
scale only axis,
xmin=1,
xmax=6,
xlabel style={font=\color{white!15!black}},
xlabel={$\text{E}_\text{b}\text{/N}_\text{0}\;\text{(dB)}$},
ymode=log,
ymin=4.9e-101,
ymax=1,
ylabel style={font=\color{white!15!black}},
ylabel={Bounds on BER},
yminorticks=true,
axis background/.style={fill=white},
xmajorgrids,
ymajorgrids,
yminorgrids,
legend style={at={(0.05,0.05)}, font=\footnotesize,anchor=south west, legend cell align=left, align=left, draw=white!15!black}
]
\addplot [color=red, dashed, line width=2.0pt]
table[row sep=crcr]{%
	0	2.59973318149547e+48\\
	0.1	1.83571276539807e+46\\
	0.2	1.18309250651194e+44\\
	0.3	6.95248008825227e+41\\
	0.4	3.72178928089043e+39\\
	0.5	1.81324043774564e+37\\
	0.6	8.03285545173413e+34\\
	0.7	3.23322387304222e+32\\
	0.8	1.1814534921197e+30\\
	0.9	3.91653867782643e+27\\
	1	1.17709687250519e+25\\
	1.1	3.20551300426543e+22\\
	1.2	7.90580696961158e+19\\
	1.3	1.76516928296729e+17\\
	1.4	356690264127020\\
	1.5	652205584454.239\\
	1.6	1079061374.72849\\
	1.7	1615546.3233292\\
	1.8	2189.36760081126\\
	1.9	2.686819851531\\
	2	0.00298803205798424\\
	2.1	3.15178109309571e-06\\
	2.2	1.34696253817189e-07\\
	2.3	1.25851306426461e-07\\
	2.4	1.19929698630987e-07\\
	2.5	1.14178958145684e-07\\
	2.6	1.08597197604576e-07\\
	2.7	1.03184615816798e-07\\
	2.8	9.79412052997195e-08\\
	2.9	9.28667485616159e-08\\
	3	8.79608167304609e-08\\
	3.1	8.3222768820297e-08\\
	3.2	7.86517516587704e-08\\
	3.3	7.4246700498039e-08\\
	3.4	7.00063403283588e-08\\
	3.5	6.59291879104377e-08\\
	3.6	6.20135545391339e-08\\
	3.7	5.82575495472697e-08\\
	3.8	5.4659084554252e-08\\
	3.9	5.12158784598434e-08\\
	4	4.79254631788606e-08\\
	4.1	4.47851901077857e-08\\
	4.2	4.17922373093195e-08\\
	4.3	3.89436173958056e-08\\
	4.4	3.62361860872561e-08\\
	4.5	3.3666651414457e-08\\
	4.6	3.12315835323765e-08\\
	4.7	2.89274251038905e-08\\
	4.8	2.67505022087345e-08\\
	4.9	2.46970357276568e-08\\
	5	2.27631531470211e-08\\
	5.1	2.09449007246872e-08\\
	5.2	1.92382559539098e-08\\
	5.3	1.7639140258329e-08\\
	5.4	1.6143431847929e-08\\
	5.5	1.47469786631738e-08\\
	5.6	1.34456113324526e-08\\
	5.7	1.22351560665255e-08\\
	5.8	1.11114474128994e-08\\
	5.9	1.00703407930305e-08\\
	6	9.10772474596058e-09\\
	6.1	8.21953280347929e-09\\
	6.2	7.40175492417926e-09\\
	6.3	6.65044841681976e-09\\
	6.4	5.96174828724651e-09\\
	6.5	5.33187694769191e-09\\
	6.6	4.7571532325753e-09\\
	6.7	4.23400067088467e-09\\
	6.8	3.7589549717927e-09\\
	6.9	3.32867068726759e-09\\
	7	2.93992702300214e-09\\
	7.1	2.5896327769091e-09\\
	7.2	2.27483039261599e-09\\
	7.3	1.9926991237303e-09\\
	7.4	1.7405573130212e-09\\
	7.5	1.51586379895955e-09\\
	7.6	1.31621847015486e-09\\
	7.7	1.13936199600778e-09\\
	7.8	9.83174769245422e-10\\
	7.9	8.456751028147e-10\\
	8	7.25016729775615e-10\\
	8.1	6.19485660271951e-10\\
	8.2	5.27496454282502e-10\\
	8.3	4.47587972609562e-10\\
	8.4	3.78418671394712e-10\\
	8.5	3.18761507335779e-10\\
	8.6	2.67498521700705e-10\\
	8.7	2.23615171201044e-10\\
	8.8	1.86194472824606e-10\\
	8.9	1.54411027876639e-10\\
	9	1.27524987801256e-10\\
	9.1	1.04876020925093e-10\\
	9.2	8.58773351715459e-11\\
	9.3	7.00098071364568e-11\\
	9.4	5.68162628027775e-11\\
	9.5	4.5895949716867e-11\\
	9.6	3.68992347693813e-11\\
	9.7	2.95225559360986e-11\\
	9.8	2.35036505521043e-11\\
	9.9	1.86170770244034e-11\\
	10	1.46700414326487e-11\\
};
\addlegendentry{BCC}

\addplot [color=blue, dashed, line width=2.0pt, mark=asterisk, mark options={solid, blue}]
table[row sep=crcr]{%
	0	5.83479519485269e+33\\
	0.5	2.69322011869566e+23\\
	1	844100668986.013\\
	1.5	0.155346755149862\\
	2	1.68690001553729e-15\\
	2.5	1.31754028706751e-30\\
	3	9.73192341646061e-47\\
	3.5	8.75084274143692e-64\\
	4	2.06553826053434e-79\\
	4.5	1.26377665436323e-88\\
	5	6.02478314057293e-99\\
	5.5	1.62474920011768e-110\\
	6	1.74310517795119e-123\\
	6.5	5.01294602962834e-138\\
	7	2.48174528965677e-154\\
	7.5	1.2868579084214e-172\\
	8	4.0022186543447e-193\\
	8.5	3.99395683639993e-216\\
	9	6.33880781040909e-242\\
	9.5	7.27891589837369e-271\\
	10	2.49907744585114e-303\\
};
\addlegendentry{HCC}

\addplot [color=blue, dashed, line width=2.0pt, mark=asterisk, mark options={solid, blue}, forget plot]
table[row sep=crcr]{%
	0	2.91739759742634e+33\\
	0.5	1.34661005934783e+23\\
	1	422050334493.007\\
	1.5	0.0776733775763748\\
	2	6.65040171986944e-13\\
	2.5	2.89607283990355e-13\\
	3	1.19660507639199e-13\\
	3.5	4.70175065748866e-14\\
	4	1.77365416759004e-14\\
	4.5	6.52847459903677e-15\\
	5	2.39004220707178e-15\\
	5.5	8.8098829173984e-16\\
	6	3.25419898999524e-16\\
	6.5	1.17852273373363e-16\\
	7	4.06311651881909e-17\\
	7.5	1.29633042514568e-17\\
	8	3.73478307339253e-18\\
	8.5	9.50498501514067e-19\\
	9	2.09133395469223e-19\\
	9.5	3.88990075009214e-20\\
	10	5.97111341950619e-21\\
};
\addplot [color=blue, dashed, line width=2.0pt, mark=asterisk, mark options={solid, blue}, forget plot]
table[row sep=crcr]{%
	0	2.91739759742634e+33\\
	0.5	1.34661005934783e+23\\
	1	422050334493.007\\
	1.5	0.0776733775763748\\
	2	6.65040171986944e-13\\
	2.5	2.89607283990355e-13\\
	3	1.19660507639199e-13\\
	3.5	4.70175065748866e-14\\
	4	1.77365416759004e-14\\
	4.5	6.52847459903677e-15\\
	5	2.39004220707178e-15\\
	5.5	8.8098829173984e-16\\
	6	3.25419898999524e-16\\
	6.5	1.17852273373363e-16\\
	7	4.06311651881909e-17\\
	7.5	1.29633042514568e-17\\
	8	3.73478307339253e-18\\
	8.5	9.50498501514067e-19\\
	9	2.09133395469223e-19\\
	9.5	3.88990075009214e-20\\
	10	5.97111341950619e-21\\
};
\addplot [color=red, dashed, line width=2.0pt, forget plot]
table[row sep=crcr]{%
	0	5.19946636299096e+48\\
	0.1	3.67142553079613e+46\\
	0.2	2.36618501302388e+44\\
	0.3	1.39049601765046e+42\\
	0.4	7.44357856178082e+39\\
	0.5	3.62648087549128e+37\\
	0.6	1.60657109034683e+35\\
	0.7	6.46644774608445e+32\\
	0.8	2.3629069842394e+30\\
	0.9	7.83307735565289e+27\\
	1	2.35419374501037e+25\\
	1.1	6.41102600853088e+22\\
	1.2	1.58116139392232e+20\\
	1.3	3.53033856593459e+17\\
	1.4	713380528254042\\
	1.5	1304411168908.47\\
	1.6	2158122749.457\\
	1.7	3231092.64665807\\
	1.8	4378.73520130662\\
	1.9	5.37363940066829\\
	2	0.00597577489336794\\
	2.1	6.0271807583264e-06\\
	2.2	5.51994256935665e-09\\
	2.3	4.59723137014595e-12\\
	2.4	3.48816172377017e-15\\
	2.5	2.41664879907572e-18\\
	2.6	1.5329847081929e-21\\
	2.7	8.9332920089934e-25\\
	2.8	4.80150693509223e-28\\
	2.9	2.39190300749717e-31\\
	3	1.11087846209965e-34\\
	3.1	4.84471163262622e-38\\
	3.2	2.00168408171307e-41\\
	3.3	7.92249008882283e-45\\
	3.4	3.04659585672334e-48\\
	3.5	1.15989672657066e-51\\
	3.6	4.55868979992445e-55\\
	3.7	5.45037222882743e-58\\
	3.8	1.69785894590017e-59\\
	3.9	7.42682258042216e-61\\
	4	3.06104082760125e-62\\
	4.1	1.17971565451355e-63\\
	4.2	4.24049803961276e-65\\
	4.3	1.41833254373475e-66\\
	4.4	4.40446093538556e-68\\
	4.5	1.26711331428014e-69\\
	4.6	3.36986070927992e-71\\
	4.7	8.26707133872978e-73\\
	4.8	1.86681918886266e-74\\
	4.9	3.87191086647739e-76\\
	5	7.35992878865898e-78\\
	5.1	1.27934637116072e-79\\
	5.2	2.0290673181818e-81\\
	5.3	2.92961648694204e-83\\
	5.4	3.84172723218442e-85\\
	5.5	4.5647929663537e-87\\
	5.6	4.90289478870584e-89\\
	5.7	4.74852597356404e-91\\
	5.8	4.13672221225056e-93\\
	5.9	3.23325178134048e-95\\
	6	2.26141129962159e-97\\
	6.1	1.41163934124018e-99\\
	6.2	7.84322448602029e-102\\
	6.3	3.86802433103734e-104\\
	6.4	1.68841035307654e-106\\
	6.5	6.50435921498645e-109\\
	6.6	2.20487634976498e-111\\
	6.7	6.55698146648514e-114\\
	6.8	1.70537914185e-116\\
	6.9	3.86689330689564e-119\\
	7	7.61944894113244e-122\\
	7.1	1.30037628848041e-124\\
	7.2	1.91571048562996e-127\\
	7.3	2.42774890030789e-130\\
	7.4	2.6372664956819e-133\\
	7.5	2.44685756451579e-136\\
	7.6	1.93178914822759e-139\\
	7.7	1.29289142348077e-142\\
	7.8	7.30689252900021e-146\\
	7.9	3.47336102735936e-149\\
	8	1.38309536042692e-152\\
	8.1	4.59448716831765e-156\\
	8.2	1.26782911289723e-159\\
	8.3	2.89358673667953e-163\\
	8.4	5.43793536159262e-167\\
	8.5	8.376801555602e-171\\
	8.6	1.05280553233334e-174\\
	8.7	1.07442484188567e-178\\
	8.8	8.86025225829176e-183\\
	8.9	5.87480397802201e-187\\
	9	3.11603784873149e-191\\
	9.1	1.3152474147846e-195\\
	9.2	4.39428524632144e-200\\
	9.3	1.15577114694776e-204\\
	9.4	2.37974190945539e-209\\
	9.5	3.81397120066802e-214\\
	9.6	4.73011668215266e-219\\
	9.7	4.51243979339962e-224\\
	9.8	3.29104506936198e-229\\
	9.9	1.82353828844933e-234\\
	10	7.62723339431275e-240\\
};
\addplot [color=blue, dashed, line width=2.0pt, mark=asterisk, mark options={solid, blue}, forget plot]
table[row sep=crcr]{%
	0	2.94686626002661e+33\\
	0.5	1.36021218115943e+23\\
	1	426313469184.855\\
	1.5	0.0784579571463949\\
	2	8.51969704816813e-16\\
	2.5	6.65424387407849e-31\\
	3	4.95595542986756e-47\\
	3.5	3.00669711153673e-54\\
	4	5.34922201174505e-60\\
	4.5	1.93287072526881e-66\\
	5	1.16703618844879e-73\\
	5.5	9.45867937507813e-82\\
	6	8.04827532774767e-91\\
	6.5	5.45638239597988e-101\\
	7	2.16265438225738e-112\\
	7.5	3.54047943875019e-125\\
	8	1.62108027911558e-139\\
	8.5	1.34033440199667e-155\\
	9	1.2249032575657e-173\\
	9.5	7.13321663833151e-194\\
	10	1.42699254658577e-216\\
};
\addplot [color=red, dashed, line width=2.0pt, forget plot]
table[row sep=crcr]{%
	0	2.6259931126217e+48\\
	0.1	1.8542553185839e+46\\
	0.2	1.19504293587064e+44\\
	0.3	7.02270715985079e+41\\
	0.4	3.75938311201052e+39\\
	0.5	1.83155599772287e+37\\
	0.6	8.11399540579206e+34\\
	0.7	3.26588270004265e+32\\
	0.8	1.19338736577748e+30\\
	0.9	3.95609967457217e+27\\
	1	1.18898673990423e+25\\
	1.1	3.23789192350044e+22\\
	1.2	7.98566360566827e+19\\
	1.3	1.78299927572454e+17\\
	1.4	360293196087900\\
	1.5	658793519650.743\\
	1.6	1089960984.57424\\
	1.7	1631864.97305963\\
	1.8	2211.48242490233\\
	1.9	2.71395929326682\\
	2	0.0030180681279636\\
	2.1	3.04403068602344e-06\\
	2.2	2.78784978250336e-09\\
	2.3	2.32183402532624e-12\\
	2.4	1.76169784028796e-15\\
	2.5	1.22052969650289e-18\\
	2.6	7.74234701107527e-22\\
	2.7	4.51176364090576e-25\\
	2.8	2.42500350257183e-28\\
	2.9	1.20803182196827e-31\\
	3	5.61049728333157e-35\\
	3.1	2.44682405688265e-38\\
	3.2	1.01095155647934e-41\\
	3.3	4.00125766548556e-45\\
	3.4	1.53868801706693e-48\\
	3.5	5.86026987996207e-52\\
	3.6	2.44359553725945e-55\\
	3.7	1.12492092002161e-57\\
	3.8	5.6534897814387e-59\\
	3.9	2.91138557959629e-60\\
	4	1.40934726140697e-61\\
	4.1	6.39391744260495e-63\\
	4.2	2.7128139239879e-64\\
	4.3	1.07420582946718e-65\\
	4.4	3.96185275265119e-67\\
	4.5	1.35828165653249e-68\\
	4.6	4.32017530957386e-70\\
	4.7	1.27224158341331e-71\\
	4.8	3.46198290681578e-73\\
	4.9	8.68738079966121e-75\\
	5	2.0061886103237e-76\\
	5.1	4.25472058156585e-78\\
	5.2	8.26931753192832e-80\\
	5.3	1.46972071113974e-81\\
	5.4	2.38350856553098e-83\\
	5.5	3.51922973637856e-85\\
	5.6	4.71998701925826e-87\\
	5.7	5.73706414878253e-89\\
	5.8	6.30473819824547e-91\\
	5.9	6.24919513811073e-93\\
	6	5.57300888296419e-95\\
	6.1	4.4603599182391e-97\\
	6.2	3.19555802130836e-99\\
	6.3	2.04398515613227e-101\\
	6.4	1.16411353997294e-103\\
	6.5	5.88716067115364e-106\\
	6.6	2.63626493971832e-108\\
	6.7	1.0423104206004e-110\\
	6.8	3.62788946415887e-113\\
	6.9	1.1082943786413e-115\\
	7	2.9625511393477e-118\\
	7.1	6.90750701923695e-121\\
	7.2	1.40031422074024e-123\\
	7.3	2.46008616373596e-126\\
	7.4	3.73280244111642e-129\\
	7.5	4.87511886732901e-132\\
	7.6	5.46100019367182e-135\\
	7.7	5.22795680094225e-138\\
	7.8	4.26152030746426e-141\\
	7.9	2.94667959478473e-144\\
	8	1.72171879059483e-147\\
	8.1	8.46718203043003e-151\\
	8.2	3.49067727938952e-154\\
	8.3	1.20138101070172e-157\\
	8.4	3.43729370070095e-161\\
	8.5	8.14027961192663e-165\\
	8.6	1.58864630414366e-168\\
	8.7	2.54339686605876e-172\\
	8.8	3.32497374567648e-176\\
	8.9	3.53257104401947e-180\\
	9	3.03539894247202e-184\\
	9.1	2.09897605345261e-188\\
	9.2	1.16214565879542e-192\\
	9.3	5.12528819194477e-197\\
	9.4	1.79090051767007e-201\\
	9.5	4.9312518068823e-206\\
	9.6	1.06403911350375e-210\\
	9.7	1.78895170309922e-215\\
	9.8	2.32996082629877e-220\\
	9.9	2.33677601749756e-225\\
	10	1.79371419822568e-230\\
};
\end{axis}

\begin{axis}[%
width=7cm,
height=6.2cm,
at={(0.788in,0.595in)},
scale only axis,
xmin=0,
xmax=1,
ymin=0,
ymax=1,
axis line style={draw=none},
ticks=none,
axis x line*=bottom,
axis y line*=left
]
\node[below right, align=left]
at (rel axis cs:0.615,0.48) {$\alpha=0.99$};
\draw[thick][->] (70,19) -- (70,13);
\draw [thick][->] (70,19) -- (85,19);
\draw [thick][->] (70,43) -- (70,35);
\draw [thick][->] (70,43) -- (80,25);
\node[below right, align=left]
at (rel axis cs:0.615,0.25) {$\alpha=0.5$};
\draw [black] (axis cs:0.712656,0.875312) ellipse [x radius=0.0197095, y radius=0.0548628];
\node[below right, align=left]
at (rel axis cs:0.55,0.83) {Union Bounds};
\end{axis}
\end{tikzpicture}%
	\caption{Expurgated union bound of HCCs and BCCs for $\alpha=0.5$ and $\alpha=0.99$, $K=512$, $R=1/3$.}
	\label{EXBERBCCHCC}
\end{figure}
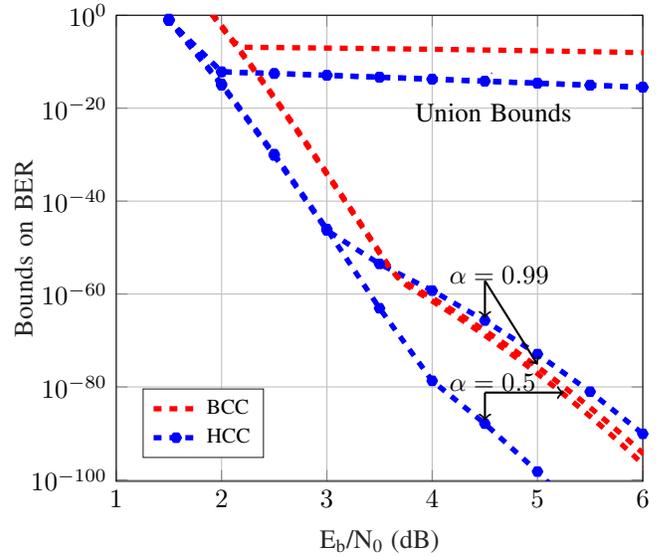

For the BCC and HCC ensembles, the gap between the expurgated bound and the union bound is very large and notable.
To investigate the influence of expurgation on the performance of these ensembles, in Fig.~\ref{EXBERBCCHCC} we provide the expurgated bound on the BER of the BCC and HCC ensembles for $\alpha=0.5$ and $\alpha=0.99$.
Note that for $\alpha=0.99$, the expurgated bounds are computed by excluding only $1\%$ of the possible codes, and these bounds are still significantly lower and steeper than the corresponding union bounds. 
For the BCC ensemble, the gap between the expurgated bounds for $\alpha=0.5$ and $\alpha=0.99$ is much smaller than that of the HCC ensemble. Therefore, for $\alpha=0.99$, the BCC ensemble has slightly steeper and lower error floor than the HCC ensemble.
The fact that changing $\alpha$ has a little impact on the expurgation of the BCC ensemble suggests that only a small fraction of the codes have poor distance properties.
This means that for a BCC with randomly selected but fixed permutations, with high probability the error floor is as steep and low as the corresponding expurgated bound for an ML decoder. 


\section{Discussion and Conclusion}\label{con}
We investigated the performance of SC-TC ensembles with finite block length in both waterfall and error floor regions.
The two primary results can be summarized as follows.
First, spatial coupling not only improves the asymptotic decoding threshold of an ensemble but also, for finite length and given latency, it improves the performance of the ensemble in the waterfall region.
Second, considering certain conditions, spatial coupling either improves or preserves the minimum distance of the ensemble. Therefore, the coupled ensembles cannot have worse error floor than the corresponding uncoupled TC ensembles.
Using this fact, instead of performing the cumbersome WEF analysis for the coupled ensemble, we derived the WEF for uncoupled ensembles. Then, based on the WEF, we computed bounds on BER performance and the minimum distance of TCs.
As the coupled ensembles have equal or larger minimum distance than the uncoupled ensembles, the computed bounds for TCs can be used to approximate the error floor of SC-TCs.
The results from the WEF analysis of TCs demonstrate very low error floors for  SCC, BCC, and HCC esnembles. Moreover, for the BCC and HCC ensembles, the minimum distance grows linearly with block length.

Comparing the thresholds of SC-TC ensembles (see Table~\ref{ThresholdsAWGN}) and the results from the WEF analysis, we observe that the ensembles with better MAP thresholds also have larger minimum distance and lower error floor. 
However, so far, only PCCs have been widely used in various standards---such as UMTS and LTE--- because of their good BP thresholds and good performance in the waterfall region.
Other TC ensembles have received much less attention for commercial use, although they have better MAP threshold and distance properties than PCCs. 
Our results confirm that the BP thresholds of these ensembles can be significantly improved by applying coupling.
Also, regarding the finite length regime, while their error floor stays at very low error probabilities, their waterfall performance gets much closer to capacity.
This brings us to the conclusion that by coupling a given ensemble with close to capacity MAP threshold and low error floor, such as SCCs, BCCs, and HCCs, the resulting ensemble is very promising and can perform close-to-capacity, yet achieving low error floor, with a low complexity iterative decoder.  

Finally, we should remark that the considered bounds estimate the error floor of an ML decoder. 
To obtain bounds on the performance of the BP decoder, more investigations on the corresponding absorbing sets \cite{Mitchell_AS2014} and pseudo-codewords \cite{RosnesPseudoweight} need to be done.
\begin{appendices}
\section{Proof of {\emph{Theorem 2}}}
We prove the theorem for the general case of tailbiting.
	\subsection{Serially Concatenated Codes} Consider the SCC and SC-SCC ensembles in Fig.~\ref{CG}(a) and (b), and assume that \[\Pi^{\text{Un}}=\Pi^{(1)}\cdot\Pi^{(2)}.\]  
	Any codeword $\bs{v}\in\mathcal{C}$ satisfies the local constraints for $t=1,\dots,L$. Therefore, at time $t$,	
	\begin{align}
	\label{SCCProof1}
	&\big(\bs{u}_t\;\; \bs{v}_t^{\text{O}}\big)\cdot \bs{H}_{\text{O}}^T=\bs{0} \; ,\\
	\label{SCCProof2}
	&\big((\tilde{\bs{v}}'_{t,0}\;\;\; \tilde{\bs{v}}'_{t-1,1}) \cdot \Pi^{(2)} \;\;\bs{v}_t^{\text{I}}\big)\cdot \bs{H}_{\text{I}}^T=\bs{0} \; ,
	\end{align}
	where $\tilde{\bs{v}}'_t=(\bs{u}_t,\bs{v}_t^{\text{O}})\cdot \Pi^{(1)}$.
	The constraints are linear and time-invariant.
	Therefore, any superposition of the vectors $\big(\bs{u}_t\;\; \bs{v}_t^{\text{O}}\big)$ and $\big((\tilde{\bs{v}}'_{t-1,1}\;\;\; \tilde{\bs{v}}'_{t,2}) \cdot \Pi^{(2)} \;\;\bs{v}_t^{\text{I}}\big)$ from different time slots $t=1,\dots,L$, satisfy \eqref{SCCProof1} and \eqref{SCCProof2}, respectively. In particular, we can consider
	\begin{align}
	\label{SCCProof3}
	\sum_{t=1}^{L}\big(\bs{u}_t\;\; \bs{v}_t^{\text{O}}\big)=\big(\sum_{t=1}^{L}\bs{u}_t\;\; \sum_{t=1}^{L}\bs{v}_t^{\text{O}}\big),
	\end{align}
	and
	\begin{align}
	&\sum_{t=1}^{L}\big((\tilde{\bs{v}}'_{t,0}\;\;\; \tilde{\bs{v}}'_{t-1,1}) \cdot \Pi^{(2)} \;\;\bs{v}_t^{\text{I}}\big)\nonumber\\
	&=\big(\sum_{t=1}^{L}(\tilde{\bs{v}}'_{t,0}\;\;\; \tilde{\bs{v}}'_{t-1,1}) \cdot \Pi^{(2)} \;\;\sum_{t=1}^{L}\bs{v}_t^{\text{I}}\big)\nonumber\\
	&=\big(\sum_{t=1}^{L}\tilde{\bs{v}}'_{t} \cdot \Pi^{(2)} \;\;\sum_{t=1}^{L}\bs{v}_t^{\text{I}}\big)\nonumber\\
	\label{SCCProof4} &=\big(\sum_{t=1}^{L}(\bs{u}_t\;\;\bs{v}_t^{\text{O}})\cdot\Pi^{(1)}\cdot\Pi^{(2)} \;\;\sum_{t=1}^{L}\bs{v}_t^{\text{I}}\big).
	\end{align}
	Let
	\[
	\tilde{\bs{u}}=\sum_{t=1}^{L}\bs{u},\;\;\;\tilde{\bs{v}}^{\text{O}}=\sum_{t=1}^{L}\bs{v}^{\text{O}},\;\;\;\tilde{\bs{v}}^{\text{I}}=\sum_{t=1}^{L}\bs{v}^{\text{I}},
	\]
	and substitute \eqref{SCCProof3} and \eqref{SCCProof4} into \eqref{SCCProof1} and \eqref{SCCProof2}, respectively. Then
	\begin{align}
	&\big(\tilde{\bs{u}}\;\; \tilde{\bs{v}}^{\text{O}}\big)\cdot \bs{H}_{\text{O}}^T=\bs{0} \; , \\
	&\big((\tilde{\bs{u}}\;\;\; \tilde{\bs{v}}^{\text{O}}) \cdot \Pi^{\text{Un}} \;\;\tilde{\bs{v}}^{\text{I}}\big)\cdot \bs{H}_{\text{I}}^T=\bs{0} \; .
	\end{align}
	Therefore, $\tilde{\bs{v}}=(\tilde{\bs{u}},\tilde{\bs{v}}^{\text{O}},\tilde{\bs{v}}^{\text{I}})$ is a codeword of the uncoupled code.
	If all nonzero elements of $\bs{v}_t$, $t=1,\dots,L$, occur at different positions, then  $w_{\text H}(\tilde{\bs{v}})=w_{\text H}(\bs{v})$. Otherwise, the overlap of the non zero elements reduces the weight of $\tilde{\bs{v}}$ and $w_{\text{H}}(\tilde{\bs{v}})< w_{\text{H}}(\bs{v})$.
	\subsection{Braided Convolutional Codes} Consider the SCC and SC-SCC ensembles in Fig.~\ref{CG}(c) and (d), and assume that $\Pi_t=\Pi$, $\Pi_t^{\text{U}}=\Pi^{\text{U}}$ and $\Pi_t^{\text{L}}=\Pi^{\text{L}}$.
	A valid code sequence of $\mathcal{C}$ has to satisfy the local
	constraints
	\begin{align}
	\begin{pmatrix}
	\bs{u}_t & \bs{v}_{t-1}^{\text{L}} \cdot \Pi_t^{\text{U}} &
	\bs{v}_{t}^{\text{U}} 
	\end{pmatrix}
	\cdot \bs{H}_{\text{U}}^T  & = \bs{0} \label{eq:coupledUpper} \; , \\
	\begin{pmatrix}
	\bs{u}_t \cdot \Pi_t & \bs{v}_{t-1}^{\text{U}} \cdot \Pi_t^{\text{L}} &
	\bs{v}_{t}^{\text{L}} 
	\end{pmatrix}
	\cdot \bs{H}_{\text{L}}^T  & = \bs{0} \label{eq:coupledLower} 
	\end{align}
	for all $t=1,\dots,L$, where $\bs{H}_{\text{U}}$ and
	$\bs{H}_{\text{L}}$ are the parity-check matrices that represent the
	constraints imposed by the trellises of the upper and lower component
	encoders, respectively.  Since these constraints are linear and time-invariant, it follows
	that any superposition of vectors
	$\bs{v}_t=(\bs{u}_t,\bs{v}_t^{\text{U}},\bs{v}_t^{\text{U}})$ from
	different time instants $t \in \{1,\dots,L\}$ will also satisfy
	\eqref{eq:coupledUpper} and \eqref{eq:coupledLower}. In particular, if
	we let
	\[
	\tilde{\bs{u}}=\sum_{t=1}^{L} \bs{u}_t \ , \quad
	\tilde{\bs{v}}^{\text{L}}=\sum_{t=1}^{L} \bs{v}_t^{\text{L}}  \ , \quad \tilde{\bs{v}}^{\text{U}}=\sum_{t=1}^{L} \bs{v}_t^{\text{U}}
	\]
	then 
	\begin{align}
	\begin{pmatrix}
	\tilde{\bs{u}} & \tilde{\bs{v}}^{\text{L}} \cdot \Pi^{\text{U}} &
	\tilde{\bs{v}}^{\text{U}} 
	\end{pmatrix}
	\cdot \bs{H}_{\text{U}}^T  & = \bs{0} \label{eq:uncoupledUpper} \; , \\
	\begin{pmatrix}
	\tilde{\bs{u}} \cdot \Pi & \tilde{\bs{v}}^{\text{U}} \cdot \Pi^{\text{L}} &
	\tilde{\bs{v}}^{\text{L}} 
	\end{pmatrix}
	\cdot \bs{H}_{\text{L}}^T  & = \bs{0} \label{eq:uncoupledLower}
	\enspace .
	\end{align}
	Here we have implicitly made use of the fact that $\bs{v}_t=\bs{0}$
	for $t<1$ and $t>L$. But now it follows from \eqref{eq:uncoupledUpper}
	and \eqref{eq:uncoupledLower} that $\tilde{\bs{v}}=(\tilde{\bs{u}},
	\tilde{\bs{v}}^{\text{U}}, \tilde{\bs{v}}^{\text{L}}) \in
	\tilde{\mathcal{C}}$, i.e., we obtain a codeword of the uncoupled
	code. If all nonzero symbols within $\bs{v}_t$ occur at
	different positions for $t=1,\dots,L$, then
	$w_{\text{H}}(\tilde{\bs{v}}) =  w_{\text{H}}({\bs{v}})$. If, on the
	other hand, the support of nonzero symbols overlaps, the weight of
	$\tilde{\bs{v}}$ is reduced accordingly and  $w_{\text{H}}(\tilde{\bs{v}}) <  w_{\text{H}}({\bs{v}})$.
%
%

The same result can be proved for HCCs by combining the proofs for PCCs and SCCs. 
\end{appendices}

\end{document}